\documentclass{elsart}
\pdfoutput=1

\usepackage[utf8]{inputenc}
\usepackage{verbatim,amsmath,amsfonts}
\usepackage{subfig}
\usepackage{appendix}
\usepackage[pdftex]{graphicx}
\usepackage{epstopdf}
\usepackage[pagewise]{lineno}
\usepackage{paralist}
\begin{document}

\begin{frontmatter}

\title{A Low Mach Number Solver: Enhancing Applicability}

\author[univie]{Natalie Happenhofer \corauthref{cor1} \thanksref{support2}}
\ead{natalie.happenhofer@univie.ac.at}
\thanks[support2]{Supported by the Austrian Science Fund (FWF), project P20973}

\corauth[cor1]{Corresponding Author}

\author[univie,mpa]{Hannes Grimm-Strele\thanksref{support2}}
\ead{hannes.grimm-strele@univie.ac.at}

\author[univie]{Friedrich Kupka \thanksref{support4}}
\ead{friedrich.kupka@univie.ac.at}
\ead[url]{http://www.mpa-garching.mpg.de/\~{}fk/}
\thanks[support4]{Supported by the Austrian Science Fund (FWF), project P21742}

\author[univie]{Bernhard L\"ow-Baselli}
\ead{bernhard.loew-baselli@univie.ac.at}

\author[univie]{Herbert Muthsam}
\ead{herbert.muthsam@univie.ac.at}
\ead[url]{http://www.univie.ac.at/acore/}

\address[univie]{University of Vienna, Faculty of Mathematics,
Nordbergstra{\ss}e 15, A-1090 Wien, Austria}

\address[mpa]{Max-Planck Institute for Astrophysics, Karl-Schwarzschild-Strasse 1, 85741 Garching, Germany}
%-----------------------------------------------------------------------------------------------------------

\begin{abstract}

In astrophysics and meteorology there exist numerous situations where flows exhibit small velocities compared to the sound speed. 
To overcome the stringent timestep restrictions posed by the predominantly used explicit methods for integration in time 
the Euler (or Navier--Stokes) equations are usually replaced by modified versions. In astrophysics this is nearly exclusively the anelastic approximation. 
Kwatra et al. \cite{kwatra} have proposed a method with favourable time-step properties integrating the \textit{original} equations (and 
thus allowing, for example, also the treatment of shocks). We describe the extension of the method to the Navier-Stokes and two-component 
equations. -- However, when applying the extended method to problems in convection and double diffusive convection (semiconvection) we ran into numerical difficulties. 
We describe our procedure for stabilizing the method. We also investigate the behaviour of Kwatra et al.'s method for very low Mach numbers 
(down to Ma $=$ 0.001) and point out its very favourable properties in this realm for situations where the explicit counterpart of this method returns 
absolutely unusable results. Furthermore, we show that the method strongly scales over 3 orders of magnitude of processor cores and is limited only by the specific network structure of the high performance computer we use. 
\end{abstract}


\end{frontmatter}%-----------------------------------------------------------------------------------------------------------

\section{Introduction}

In astrophysics, meteorology and many other fields of physical and engineering sciences problems are studied which are characterized by vastly different timescales. In fluid 
flow the smallest time scales are often set by diffusive processes, whereas the solution changes only on the much larger time scales set by the flow velocity 
or the sound speed. It is then general practice to make use of explicit time integration for the hyperbolic terms and to treat the diffusive terms implicitly. 
The latter task is facilitated by the fact that the resulting system of discretized equations is often linear and positive definite so that effective solution 
methods are available.

Frequently, however, the different scales originate from the macroscopic flow speed (small) and the sound speed (large). In addition, 
the sound waves may be known to be of little physical significance. Such a situation is ubiquitous in meteorology and geophysics (convection in 
the earth mantle). In stellar physics, these premises typically hold true for convection in the interior of stars, for example in the solar convection 
zone except for its outermost parts. Since an implicit time-advancement of the hyperbolic part is usually impractical due to, among others, a lack of efficient 
solvers for these terms, the usual solution strategy here is to eliminate the sound waves analytically and to discretize the modified equations only. A widely 
(in stellar physics: nearly exclusively) used way to accomplish that is to resort to the anelastic approximation \cite{oguraphillips} and its many variants 
or extensions. In meteorology, the use of so-called primitive equations is also widespread (cf.,\cite{arakon},e.g.). 

In \cite{kwatra} N. Kwatra et al. propose a semi--implicit method to solve Euler's equations numerically. From a physical point of view the 
advantage of this approach is that the method avoids any simplifications on the side of the basic equations themselves. As a consequence of that 
it is possible to treat physical phenomena for which the anelastic methods are neither designed nor appropriate, e.g. shocks. Indeed, in \cite{kwatra} a 
number of shock tube problems are satisfactorily treated as test cases. Furthermore, it is outlined that this method has favourable properties in the low Mach number regime. On the numerical side, the implicit part consists of the solution of an equation 
of a generalized Poisson equation only (equation (\ref{eq:GeneralizedPoisson}) in the present paper). For such equations, efficient solvers are quite 
readily available. In \cite{kwatra}, the method was developed in the framework of ENO type numerics.

These properties made the method an interesting candidate for inclusion in our ANTARES code \cite{antares} which is designed to cope with 
flows encountered in stellar physics ranging from low to high Mach numbers, including shocks. The original development in terms of ENO 
methodology complied well with the fact that ANTARES makes use of this methodology also in the other methods included for spatial discretization.

While after a few adaptions the Kwatra method performed well in the usual shock tube tests we performed, difficulties arose when applying 
its straight forward extension to the problem of semiconvection in stellar physics (double--diffusive convection; diffusive convection in oceanography) in the same way we 
had, earlier on, successfully  tackled similar problems with standard ENO methods. In semiconvection possibly slow (in terms of sound speed) convective 
motions take place. Already they enforce the calculations to span a large time interval in order to arrive at a relaxed, statistically steady state. 
In addition, between zones of ordinary convection thin, more or less horizontal sheets are situated where the transport of helium in the star (salt in 
the ocean) is provided by diffusive processes only, which may easily make for even longer time scales \cite{zaussispruit}. Experience thus showed that 
the successful mastering of the shock tube problems as usually applied for validating numerical discretizations does not guarantee stability for long 
periods of time. The extended version of Kwatra's method, which basically consisted  of adding the viscous terms to the explicit part of the method did not yield stable results. Indeed, severe instabilities appeared in the course of time which rendered the calculations useless. This was quite surprising since the explicit ENO-based method employing the same discretization led to satisfyingly stable results. It was therefore warranted to stabilize the original method so as to achieve stability even for long-term integrations.

The plan of the paper is the following. We first give a short description of the basics of Kwatra et al.'s method. Subsequently, we derive a pressure 
evolution equation which, unlike the original equation, takes also dissipative terms into account (radiative transfer in the diffusion approximation as 
adequate for stellar interiors). We also consider the case of two-component flows (semi- or thermohaline convection). We then describe the difficulties 
which resulted when applying this method to the semiconvection problem. Next we turn to the enhancement of stability which solved the numerical problems 
and discuss a few simulations. 

Subsequently, we show that this method scales strongly over 3 orders of magnitude of processor cores and is limited only by the specific network structure of the high performance method we use. This makes it suitable for high performance computing despite the necessity of solving an elliptic equation each timestep. 

Furthermore, we validate the method of Kwatra et al., by testing it in the high Mach number regime as well as its performance in low Mach number flows (down to Ma$=$0.001). In the latter regime, Kwatra's method shows very satisfying behaviour which contrasts grossly with the completely useless results of the standard explicit method.

We conclude that, given the beneficial properties of this method, it poses a real alternative to the anelastic or Boussinesq approximations. 

%-----------------------------------------------------------------------------------------------------------

\section{Numerical method} \label{sec:numericalmethod}

We model fluid flow by the set of compressible Navier-Stokes equations (NSE). A detailed derivation is found in \cite{hillebrandtkupkaLecPhys} and \cite{lanlif63}. For better legibility, 
all derivations are presented in one spatial dimension. Hence, $\partial / \partial x$ is, in this case, the one-dimensional divergence operator $\nabla$. The generalization to higher dimensions is easily done.  

\vspace{1em}

The compressible Navier-Stokes equations for a chemically homogeneous flow read
\begin{equation}
\frac{\partial}{\partial t}
\left(\begin{array}[3]{c}
\rho \\
\rho u \\
e \\
\end{array}\right)
+
\frac{\partial}{\partial x}
\left(\begin{array}[3]{c}
\rho u \\
\rho u \cdot u + P - \tau\\
e u  + P \cdot u - u \cdot \tau - F_{\mathrm{rad}}\\
\end{array}\right) 
= 
\left(\begin{array}[3]{c}
0\\
\rho g \\
\rho g u \\
\end{array}\right) \label{eq:navstoonecom}
\end{equation}

The state variables in the Navier-Stokes equations depend in principle on the spatial variable $x$ and time $t$. The explicit dependencies are stated in 
Table \ref{Variables}. In general, the radiative source term $Q_{\mathrm{rad}} = \nabla \cdot F_{\mathrm{rad}}$ is determined as the stationary 
limit of the radiative transfer equation (see \cite{mihalas84} for discussion and further references). 
However, we will only consider settings taking place in stellar regions which are optically thick, so 
this quantity can accurately be obtained by means of the diffusion approximation for radiative transfer,
$Q_{\mathrm{rad}} = \nabla \cdot F_{\mathrm{rad}} = -\nabla \cdot (K \nabla T)$. 

\begin{table} 
 
 \centering
 \begin{tabular}{|l|l|}
\hline
 $\rho = \rho(x,t)$ & gas density \\
 $u=u(x,t)$ & flow velocity \\
 $\rho u$   & momentum density \\
 $ P = P(x,t) $ & gas pressure \\
 $ c = c(x,t) $ & concentration of second species \\
 $ g $ & gravitational acceleration \\ 
 $ \tau $ & viscous stress tensor for zero bulk viscosity \\
 $ \eta $ & dynamic viscosity, appears in the definition of $\tau$ \\
 $ T=T(x,t)$ & temperature \\
 $ K = K(\rho,T,c)$ & radiative conductivity \\
 $ \kappa_{\mathrm{T}} = K/(c_{\mathrm{P}} \rho) = \kappa(T,\rho) $ & radiative diffusivity \\
 $e = e(x,t)=e_{\mathrm{int}}(x,t) + e_{\mathrm{kin}}(x,t) $ & total energy density, sum of internal and kinetic energy densities\\
 $\gamma$ & adiabatic index ($\gamma = 5/3$) for an ideal gas \\
 $\kappa_{\mathrm{c}} = \kappa_{\mathrm{c}}(\rho,T,c) $ & diffusion coefficient for species c \\
\hline
 \end{tabular}
\caption{Variables and parameters appearing in the model equations}
 \label{Variables}
\end{table}

\vspace{1em}

We now introduce our extended version of the semi-implicit numerical method of N. Kwatra et al. \cite{kwatra}. To this end, we split the flux term 
in an advective part, a non-advective part, and a viscous part,

\begin{equation*}
F_{\mathrm{ad}}(U) = 
\left(\begin{array}[3]{c}
\rho u \\
\rho u^2  \\
e u   \\
\end{array}\right)
\text{, }
F_{\mathrm{nonad}}(U) =
\left(\begin{array}[3]{c}
0 \\
P \\
P \cdot u \\
\end{array}\right)
\text{, }
F_{\mathrm{vis}}(U) = -
\left(\begin{array}[3]{c}
0 \\
\tau \\
u \cdot \tau + K \nabla T \\
\end{array}\right) \label{eq:fluxsplitting}
\end{equation*}

The advective flux $F_{\mathrm{ad}}$ is calculated using the weighted essentially non-oscillatory (WENO) method proposed in \cite{weno}. This approach 
takes the computational values provided as cell averages $U_i =\frac{1}{h} \int\limits_{x_{i-\frac{1}{2}}}^{x_{i+\frac{1}{2}}}{v(\zeta) \, d\zeta} $ 
and reconstructs the numerical fluxes at the cell interfaces by the means of an upwind biased interpolation operator. 
To ensure that every conservative variable is upwinded according to its characteristic speed, which is given by the corresponding eigenvalue of 
the Jacobian $DF_{\mathrm{ad}}$, the upwinding is done in the eigensystem where the equations decouple. Due to the flux splitting approach 
the Jacobian of the advective flux $F_{\mathrm{ad}}$ reads

\begin{equation}
DF_{\mathrm{ad}}(U) = 
 \left(\begin{array}[3]{ccc}
0 & 1 & 0 \\
-u^2 & 2u & 0 \\
-\frac{e u}{\rho} & \frac{e}{\rho} & u \\
\end{array}\right)
\end{equation}

Since all eigenvalues of $DF_{\mathrm{ad}}$ are equal to u, we make use of the characteristic projection method (CPM) \cite{cpm} and apply componentwise 
upwinding to $F_{\mathrm{ad}}(U)$. However, Kwatra et al. found in \cite{kwatra} that a slight modification of the advective flux function $F_{\mathrm{ad}}$ leads to less spurious osciallatory behaviour of the ENO scheme. They propose to replace the advection velocities with the MAC grid value defined at the point in question, i.e. replace $F_{\mathrm{ad}} = (\rho_j u_j, \rho_j u_j^2, e_j u_j)$ with $\tilde{F}_{\mathrm{ad}}=(\rho_j u_{i+1/2},\rho_j u_j u_{i+1/2}, e_j u_{i+1/2})$ with $u_{i+1/2} = \frac{(\rho u)_i + (\rho u)_{i+1}}{\rho_i + \rho_{i+1}}$ to construct the flux at $x_{i+1/2}$. They call this approach \textit{modified ENO (MENO)}. 

\vspace{1em}

The viscous part $F_{\mathrm{vis}}(U)$ is calculated using a fourth order finite difference scheme presented in Section \ref{sec:dissipativestencil}.

%\vspace{1em}
Thus, we update the advective and the viscous part and add the vector of body forces to obtain an intermediate value of the conservative 
variables $\rho^*$, $(\rho u)^{*}$ and $e^*$. Since the pressure does not affect the density $\rho$, we have $\rho^* = \rho^{n+1}$. Here and in the sequel superscripts like $n$ denote the timestep.  

\vspace{1em}

As outlined in \cite{kwatra}, the non-advective momentum and energy updates are

\begin{eqnarray}
\frac{(\rho u)^{n+1} - (\rho u)^*}{\Delta t} = - \nabla P \label{eq:momupdate} \\
\frac{e^{n+1} - e^*}{\Delta t} = - \nabla \cdot (Pu) \label{eq:enupdate}
\end{eqnarray}

Dividing Eq. (\ref{eq:momupdate}) by $\rho^{n+1}$ and taking its divergence we obtain 

\begin{equation}
\nabla \cdot u^{n+1} = \nabla \cdot u^* - \Delta t \nabla \cdot \left(\frac{\nabla P}{\rho^{n+1}}\right) \label{eq:divu}
\end{equation}

Similar to \cite{kwatra}, we use the pressure evolution equation for the Navier-Stokes equations derived below in Section \ref{sec:pressureevolution},

\begin{equation}
\frac{\partial P}{\partial t} + u \cdot \nabla P = -(\nabla \cdot u) \rho c_s^2 - \frac{2}{3\rho} \left( u \cdot \nabla \tau - \nabla \cdot (u \cdot \tau) + \nabla \cdot (K \nabla T) \right) \label{eq:pressureevolution}
\end{equation}

We take $\nabla \cdot u$ to be at time $n+1$ through the timestep and substitute into (\ref{eq:divu}) to obtain

\begin{equation}
\frac{\partial P}{\partial t} + u \cdot \nabla P = - \rho c_s^2 \nabla \cdot u^* + \rho c_s^2 \Delta t \nabla \cdot \left(\frac{\nabla P}{\rho^{n+1}}\right) - 
\frac{2}{3\rho} \left(u \cdot \nabla \tau - \nabla \cdot (u \cdot \tau) + \nabla \cdot (K \nabla T) \right) \label{eq:pressev1}
\end{equation}

This is an advection-diffusion equation with a source term. We discretize the advective term and the source term using an explicit Euler-forward 
timestep and define the pressure in the diffusion term $\nabla \cdot (\nabla P / \rho^{n+1})$ implicitly at $t^{n+1}$. Apart from $u^*$, the velocity is taken at the old timestep, $u^n$. 
This leads to a general elliptic equation 

\begin{equation}
\tilde{c}P^{n+1} - \nabla \cdot (\kappa \nabla P^{n+1}) = f \label{eq:Poiss}
\end{equation}

where

\begin{eqnarray}
\tilde{c} &=& \frac{1}{\Delta t^2 \rho^n c_s^2} \label{eq:cpoiss}\\
\kappa &=& \frac{1}{\rho^{n+1}} \label{eq:kappapoiss} \\
f &=& \frac{P^n - \Delta t\; u^n \cdot \nabla P^n}{\Delta t^2 \rho^n (c_s^n)^2} - \frac{1}{\Delta t} (\nabla \cdot u^*) \notag \\
& & -  \frac{2}{3 \Delta t\; (\rho^n)^{2} (c_s^n)^2} (u^n \cdot \nabla \cdot \tau^n - \nabla \cdot (u^n \cdot \tau^n) + \nabla \cdot(K \nabla T^n)) \label{eq:fpoiss}
\end{eqnarray}

As it is common in projection-like methods, we prescribe von Neumann boundary conditions, 

\begin{equation}
\frac{\partial P^{n+1}}{\partial \vec{n}}\mid_{\partial \Omega} = 0. \label{eq:vonNeumannBC}
\end{equation}

$\vec{n}$ denotes the outward pointing unit normal. 

\vspace{1em}

Taking over the formalism of the dual cell of \cite{kwatra} we interpolate the pressure to the cell interfaces via

\begin{equation}
P_{i+\frac{1}{2}} = \frac{P_{i+1} \rho_{i} + P_{i} \rho_{i+1}}{\rho_i + \rho_{i+1}} \label{eq:pbnd}
\end{equation}

and update equations (\ref{eq:momupdate}) and (\ref{eq:enupdate}) using a central difference quotient. 

However, the pressure P obtained form equation (\ref{eq:Poiss}) is only an accurate prediction of the thermodynamical pressure at time $n+1$. Therefore, after the completion of the momentum and energy update, the predicted pressure is discarded and the pressure at the new timestep is determined from the equation of state. This ensures that the pressure is always perfectly consistent with density and temperature values. 

%-----------------------------------------------------------------------------------------------------------

\subsection{The Pressure Evolution Equation for a general EOS} \label{sec:pressureevolution}
Basically, the derivation of the pressure evolution equation follows the lines of \cite{multispecies}, but instead of the Euler equations, we use 
the NSE. Although we start out using a general equation of state (EOS), the final shape of the pressure evolution equation depends on the 
equation of state considered, which is in our case the equation of state for ideal gas. However, a corresponding pressure evolution 
equation may be derived for any EOS. 

We consider the general equation of state $P = P(\rho,\varepsilon)$ where $\varepsilon = e_{\mathrm{int}}/\rho$. Taking the total derivative we have

\begin{equation}
\frac{DP}{Dt} = \left( \frac{\partial P}{\partial \rho}\right)_{\varepsilon} \frac{D\rho}{Dt} + \left(\frac{\partial P}{\partial \varepsilon}\right)_{\rho} 
\frac{D \varepsilon}{Dt} \label{eq:totderivP}
\end{equation}

As it is common in thermodynamical notation, the subscripts in equation (\ref{eq:totderivP}) denote the thermodynamic state variable which is held constant at the evaluation of the differential expression. 

\vspace{1em}

Analysis of the Navier-Stokes equations shows that

\begin{eqnarray}
\frac{D\rho}{Dt} &=& - \rho \nabla \cdot u \\
\frac{D\varepsilon}{Dt} &=& \frac{-u \cdot (\nabla \cdot \tau) - P (\nabla \cdot u) + \nabla \cdot (u \cdot \tau) - \nabla \cdot ( K \nabla T)}{\rho} 
\end{eqnarray}

Using this and the general definition of the sound speed, $c_{\mathrm{s}}^2 := \left( \frac{\partial P}{\partial \rho}\right)_{S}= \left( \frac{\partial P}{\partial \rho}\right)_{\varepsilon} + 
\frac{P}{\rho^2} \left(\frac{\partial P}{\partial \varepsilon} \right)_{\rho}$, where S is the specific entropy, we arrive at

\begin{equation}
\frac{\partial P}{\partial t} + u \cdot \nabla P = -(\nabla \cdot u) \rho c_{\mathrm{s}}^2 - \left( \frac{\partial P}{\partial \varepsilon} \right)_{\rho} \frac{1}{\rho^2} (u \cdot \nabla \tau - \nabla \cdot (u \cdot \tau) + \nabla \cdot ( K \nabla T)) 
\end{equation}

The partial derivative $\left(\frac{\partial P}{\partial \varepsilon}\right)_{\rho}$ is calculated using the equation of state. In case of ideal gas the EOS 
relating the pressure $P$ to the internal energy $\varepsilon$ reads

\begin{equation*} 
P = \frac{2}{3}\varepsilon \rho \text{ , so naturally, we have } \quad \left(\frac{\partial P}{\partial \varepsilon}\right)_{\rho} = \frac{2}{3} \rho
\end{equation*}

Thus we arrive at equation (\ref{eq:pressureevolution}).
%-----------------------------------------------------------------------------------------------------------

\subsection{Extension to a two-component flow}
To model the behaviour of a two component fluid we add a concentration equation to the Navier-Stokes equations:
\begin{equation}
\frac{\partial}{\partial t}(\rho c) + \frac{\partial}{\partial x} (\rho c u) = \nabla \cdot (\rho \kappa_{\mathrm{c}} \nabla c) \label{eq:conc}
\end{equation}

Splitting the fluxes as before leads to 
\begin{equation*}
F_{\mathrm{ad2}}(U) = 
\left(\begin{array}[4]{c}
\rho u \\
\rho c u \\
\rho u^2  \\
e u   \\
\end{array}\right)
\text{, }
F_{\mathrm{nonad2}}(U) =
\left(\begin{array}[4]{c}
0 \\
0 \\
P \\
P \cdot u \\
\end{array}\right)
\text{, }
F_{\mathrm{vis2}}(U) = -
\left(\begin{array}[4]{c}
0 \\
\rho \kappa_{\mathrm{c}} \nabla c \\
\tau \\
u \cdot \tau + K \nabla T \\
\end{array}\right)
\end{equation*}

and since all the eigenvalues of

\begin{equation}
DF_{\mathrm{ad2}}(U) = 
 \left(\begin{array}[4]{cccc}
0 & 0 & 1 & 0 \\
-cu & u & c & 0 \\
-u^2 & 0 &2u & 0 \\
-\frac{e u}{\rho} & 0 & \frac{e}{\rho} & u \\
\end{array}\right)
\end{equation}

equal $u$, we apply the MENO algorithm to $F_{\mathrm{ad2}}(U)$ and evaluate $F_{\mathrm{vis2}}$ as outlined for the one-component case.

\vspace{1em}

Similar to the continuity equation, equation (\ref{eq:conc}) is unaffected by the pressure, so $(\rho c)^{*} = (\rho c)^{n+1}$. 

\vspace{1em}

Basically, the momentum and energy updates (\ref{eq:momupdate}), (\ref{eq:enupdate}) are carried out as before, but since the pressure depends on the mass 
fraction $c$, the pressure evolution equation (\ref{eq:pressureevolution}) takes a slightly different appearance.

Since $P=P(\rho,\varepsilon,c)$, equation (\ref{eq:totderivP}) reads
\begin{equation}
\frac{DP}{Dt} = \left(\frac{\partial P}{\partial \rho} \right)_{\varepsilon,c} \frac{D\rho}{Dt} + \left(\frac{\partial P}{\partial \varepsilon}\right)_{\rho,c} 
\frac{D\varepsilon}{Dt} + \left(\frac{\partial P}{\partial c}\right)_{\rho,\varepsilon} \frac{Dc}{Dt} .
\end{equation}  

Enforcing mass conservation, analysis of equation (\ref{eq:conc}) shows that

\begin{equation}
\frac{Dc}{Dt} =\frac{\nabla \cdot (\rho \kappa_{\mathrm{c}} \nabla c)}{\rho} .
\end{equation}

We use the equation of state for a perfect gas

\begin{equation}
P = \frac{R_{\mathrm{gas}}\rho T}{\mu} \label{eq:idealEOS}
\end{equation}

where $\mu$ denotes the mean molecular weight of our compound. It is related to the mass fraction $c$ by $\mu = \frac{1}{c\;/\mathrm{atm}_1 + (1-c)\; /\mathrm{atm}_2}$. $\mathrm{atm}_1$ 
and $\mathrm{atm}_2$ refer to the atomic weight of the constituents of the fluid. In our case, we usually model a mixture of hydrogen and helium. 
$c$ denotes the mass fraction of helium. Therefore, we have $\mathrm{atm}_1 = 4$ being the atomic weight of helium whereas $\mathrm{atm}_2=1$ is the atomic weight of hydrogen (in a
non-ionized state). It is straightforward to include electron pressure in the EOS for a partially or a fully ionized gas.
Thus, we arrive at

\begin{equation}
\left(\frac{\partial P}{\partial c}\right)_{\rho,\varepsilon} = -\frac{3}{4} R_{\mathrm{gas}} \rho T
\end{equation}

Whence the pressure evolution equation (\ref{eq:pressureevolution}) reads

\begin{equation}
\frac{\partial P}{\partial t} + u \cdot \nabla P = -(\nabla \cdot u) \rho c_{\mathrm{s}}^2 - \frac{2}{3\rho} (u \cdot \nabla \tau - \nabla \cdot (u \cdot \tau) + \nabla \cdot (K \nabla T)) - \frac{3}{4} R_{\mathrm{gas}} T \nabla \cdot (\rho \kappa_{\mathrm{c}} \nabla c)
\end{equation}

We can thus use the discretization technique employed in equations (\ref{eq:divu}) and (\ref{eq:pressev1}) to proceed from timestep $n$ to $n+1$ and rewrite equation (\ref{eq:fpoiss}) such that the right hand side becomes

\begin{eqnarray}
f &=& \frac{P^n - \Delta t\; u^n \cdot \nabla P^n}{\Delta t^2 \rho^n (c_{\mathrm{s}}^n)^2} \nonumber \\
 &-& \frac{1}{\Delta t} (\nabla \cdot u^*) - \frac{2}{3 \Delta t (\rho^n)^2 (c_{\mathrm{s}}^n)^2} (u^n \cdot \nabla \cdot \tau^n - \nabla \cdot (u^n \cdot \tau^n) 
+ \nabla \cdot (K \nabla T^n)) \nonumber \\ 
&-& \frac{3}{4 \rho^n \Delta t (c_{\mathrm{s}}^n)^2} R_{\mathrm{gas}} T^n \nabla \cdot (\rho^n \kappa_{\mathrm{c}} \nabla c^n) \label{eq:fpoiss2}
\end{eqnarray}
%-----------------------------------------------------------------------------------------------------------
\subsection{The Algorithm}

In the end, the fractional step algorithm presented in the previous sections consists of the following steps:

\begin{enumerate}[(i)]

\item We start at timestep $n$ with values $\rho^n,(c)^n (\rho u)^n, (e)^n, P^n$.  We calculate the velocity at the cell interfaces via
   \begin{equation}
   u_{i+1/2} = \frac{(\rho u)_i + (\rho u)_{i+1}}{\rho_i + \rho_{i+1}} \label{eq:step1}
   \end{equation}
   and determine the modified flux function $\tilde{F}_{\mathrm{ad}\; j} = (\rho_j u_{i+1/2},\rho_j c_j u_{i+1/2}, \rho_j u_j u_{i+1/2}, e_j u_{i+1/2})^T$. 

\vspace{1em}

\item We determine $\frac{\partial}{\partial x} \tilde{F}_{\mathrm{ad}}$ using the CPM procedure. $\frac{\partial}{\partial x} F_{\mathrm{vis}}$ is calculated using a finite-difference approach.

\vspace{1em}

\item We determine the density at the new timestep as well as intermediate values for momentum and total energy via 
  \begin{equation}
  \left(\begin{array}[4]{c}
\rho^{n+1} \\
(\rho c)^{n+1} \\
(\rho u)^* \\
e^* \\
\end{array}\right) =
\left(\begin{array}[4]{c}
\rho^{n} \\
(\rho c)^n \\
(\rho u)^n \\
e^n \\
\end{array}\right) -
 \Delta t \left(\frac{\partial}{\partial x} \tilde{F}_{\mathrm{ad}} + \frac{\partial}{\partial x} F_{\mathrm{vis}}\right)
  \end{equation}
  
\vspace{1em}

\item Using equation (\ref{eq:pbnd}), we interpolate $P^n$ to the cell interfaces and apply a central difference quotient to calculate $\nabla P^n$,
   \begin{equation}
   (\nabla P^n)_i = \frac{P^n_{i+1/2} - P^n_{i-1/2}}{\Delta x} \label{eq:nablaP}
   \end{equation} 
   Similarly, $\nabla \cdot u^*$ is determined by interpolating $(\rho u)^*/\rho^{n+1} = u^*$ to the cell interfaces as in equation (\ref{eq:step1}) and calculate
   \begin{equation}
   (\nabla u^*)_i = \frac{u^*_{i+1/2} - u^*_{i-1/2}}{\Delta x}
   \end{equation}

\vspace{1em}

\item We now determine the coefficient functions $\tilde{c}(x)$ and $\kappa(x)$ according to equations (\ref{eq:cpoiss}) and (\ref{eq:kappapoiss}). The right hand side $f(x)$ is calculated according to (\ref{eq:fpoiss}) or (\ref{eq:fpoiss2}), respectively. 
  
\vspace{1em}

\item We solve the generalized Poisson equation (\ref{eq:GeneralizedPoisson}) prescribing von Neumann boundary conditions (\ref{eq:vonNeumannBC}) and obtain the predicted pressure $P^{n+1}_{\mathrm{pred}}$. 

\vspace{1em}

\item We determine $\nabla P^{n+1}_{\mathrm{pred}}$ analogously to $\nabla P^n$ in equation (\ref{eq:nablaP}) in step (iv) and perform the momentum update as prescribed by equation (\ref{eq:momupdate}).
   
\vspace{1em}

\item Following the suggestion of N. Kwatra et al. in \cite{kwatra}, the cell face velocity $u^*_{i+1/2}$ is updated separately to $u^{n+1}_{i+1/2}$ via 
   \begin{equation}
    u^{n+1}_{i+1/2} = u^*_{i+1/2} - \Delta t \frac{P^{n+1}_{\mathrm{pred}\;i+1} - P^{n+1}_{\mathrm{pred}\;i}}{\Delta x} \frac{2}{\rho^{n+1}_{i+1} + \rho^{n+1}_{i}}
   \end{equation}

\vspace{1em}

\item We update the total energy $e$ as prescribed in equation (\ref{eq:enupdate}), 
   \begin{equation}
   \frac{e^{n+1} - e^*}{\Delta t} = - \nabla \cdot (P_{\mathrm{pred}}u)^{n+1} = - P^{n+1}_{\mathrm{pred}} \nabla u^{n+1} - u^{n+1} \nabla P^{n+1}_{\mathrm{pred}}
   \end{equation}
   As usual, the gradients are determined via the central difference quotient using the interpolated values at the cell interfaces. 

\vspace{1em}

\item We determine the EOS pressure at time $n+1$ and set $P^{n+1} = P^{n+1}_{\mathrm{EOS}}$. 

\end{enumerate}
%-----------------------------------------------------------------------------------------------------------
\subsection{A Solver for the Generalized Poisson Equation}\label{sec:Poisson}
The fractional step method presented above relies on the prediction of the pressure at the next time step. This prediction is obtained by evaluating a Poisson type equation,

\begin{equation}
-\nabla \cdot (\kappa(x) \nabla \Phi(x)) + \tilde{c}(x)\Phi(x) = f(x) . \label{eq:GeneralizedPoisson}
\end{equation}

It is crucial for the performance of the overall method to have an efficient numerical solver for equations of this type at hand. To fit into the existing ANTARES framework this solver is required to run in parallel and scale efficiently.   

\vspace{1em}

From an analytical point of view existence and uniqueness of a solution to equation (\ref{eq:GeneralizedPoisson}) is guaranteed, if the coefficient 
functions $\kappa(x)$ and $\tilde{c}(x)$ are strictly positive and bounded over the entire domain. Due to (\ref{eq:cpoiss}) and (\ref{eq:kappapoiss}) this can be ensured for any physically meaningful problem where $\rho > 0$. The conditions on $\kappa(x)$ and $\tilde{c}(x)$ ensure the ellipticity and coercivity of the linear operator $L=-\nabla\cdot (\kappa(x) \nabla) + \tilde{c}(x)$ and therefore, we may invoke the theorem of Lax--Milgram which guarantees the existence of a unique weak solution of (\ref{eq:GeneralizedPoisson}) for any square--integrable $f(x)$. A general theoretical treatment of second order elliptic partial differential equations is found in \cite{evans}, whereas in \cite{HGSPoisson} the special case of (\ref{eq:GeneralizedPoisson}) is analysed. 

Various techniques are available to discretize equation (\ref{eq:GeneralizedPoisson}). The discretization process
transforms the partial differential equation into a system of linear equations. The solution of the linear system is the numerical approximation to 
the analytical solution of (\ref{eq:GeneralizedPoisson}).

For discretization we preferred a finite element ansatz over a finite difference scheme, since the finite element technique leads to a symmetric and 
positive definite system matrix. A solution of the resulting linear system is determined by means of the preconditioned Conjugate Gradient algorithm. 
As a preconditioner we use the incomplete Cholesky decomposition with fill--in. A detailed description of the methods is found in \cite{HGSPoisson}.

%To obtain a numerical solution of equation (\ref{eq:GeneralizedPoisson}) with given boundary conditions, it must be transformed into %a linear system. Since the finite difference approach leads to a nonsymmetric system matrix, we preferred a finite element ansatz. 
%The resulting linear system is solved by the use of the preconditioned Conjugate Gradient Algorithm. As a preconditioner we use the %Incomplete 
%Cholesky Decomposition with fill-in. For parallelization, we employ the Schur Complement Algorithm. 
%

\subsubsection{Scalability of the Poisson Solver} 

To test the scaling capability of our solver we solve the two--dimensional generalized Poisson equation 
(\ref{eq:GeneralizedPoisson}) with $\kappa(x), \tilde{c}(x), f(x)$ as specified in (\ref{eq:cpoiss})--(\ref{eq:fpoiss}) on a rectangular domain 
discretized with $800 \times 800$ grid points and employ a varying number of computational cores. The results are given in Table \ref{scaling Poisson}.

We point out that the run on a single core is algorithmically different, since in this case the generalized Poisson equation is solved using (only) the 
Conjugate Gradient algorithm. The Schur complement algorithm is first required solving the generalized Poisson equation on two cores and induces considerable 
overhead. The wallclock times indicate that this overhead slows down the calculation by a factor of 12, i.e. employing the Schur complement method and solving 
the generalized Poisson equation on 12 cores takes as long as solving it on a single core employing just the Conjugate Gradient algorithm. Therefore, in our 
test the first real speedup is achieved employing 16 cores. However, the results also show that the wallclock time decreases linearly beyond that. 
Hence, except for a constant factor, the Schur complement method scales optimally over three orders of magnitude in number of cores.  

\begin{table}[ht!]
 \centering
 \begin{tabular}{|l|l||l|l|}
\hline
  \# cores & time in s & \# cores & time in s \\\hline
  \hline
  1 &  1.000 & 64 & 0.168  \\ \hline
  2 &  6.033 & 128 &  0.096 \\\hline
  4 &  3.109 & 256 &  0.043 \\\hline
  8 &  1.361 & 512 & 0.021 \\\hline
  16 & 0.695 & 1024 & 0.016 \\\hline
  32 &  0.345 & & \\
  \hline
 \end{tabular}
\caption{Poisson solver scaling test: calculated at the IBM Power6 575 System at RZG. Due to the considerable overhead induced by the Schur complement method the first real speedup compared to the (algorithmically different) single--core run is achieved employing 16 cores. }
 \label{scaling Poisson}
\end{table}

\subsubsection{Stability of the Poisson Solver} \label{sec:StabilityPoisson}

In this test, we solve equation (\ref{eq:GeneralizedPoisson}) with $\kappa(x,y)=1$, $c(x,y)=1$ and $f(x,y)=0$ and apply Neumann boundary conditions in x--direction and periodic boundary conditions in y--direction. The analytical solution of 
  
%$\kappa(x,y)=c(x,y)=1$ and the right-hand side of REFERENZ AUF GPE is perturbed with random numbers of dimension $10^{-2}$ to get a new right-hand side %$\tilde{f}$. The unperturbed problem

\begin{align*}
  - \Delta \phi + \phi & = 0, \\
  \frac{\partial \phi}{\partial \vec{n}}|_{x=0} = -\mathrm{e}^0, & \qquad \frac{\partial \phi}{\partial \vec{n}}|_{x=1} = \mathrm{e}^1
\end{align*}

is $\phi(x,y) = \mathrm{e}^{x},\ x,y \in \left[0,1\right]$. 

To ascertain whether our solver is capable of producing smooth results even if slight oscillations are present in the coefficient functions we introduce random perturbations of the order $10^{-2}$ in $\kappa, \tilde{c}$, and $f$. 

\begin{figure}[htbp]

%       \begin{minipage}[t]{0.475\linewidth}
%         \centering
%         \includegraphics[scale=0.25]{noise.jpg}
%         \caption{The random perturbation $\tilde{f}$.}\label{noise}
%       \end{minipage}
%       \hfill
%       \begin{minipage}[t]{0.475\linewidth}
         \centering
         \includegraphics[scale=0.5]{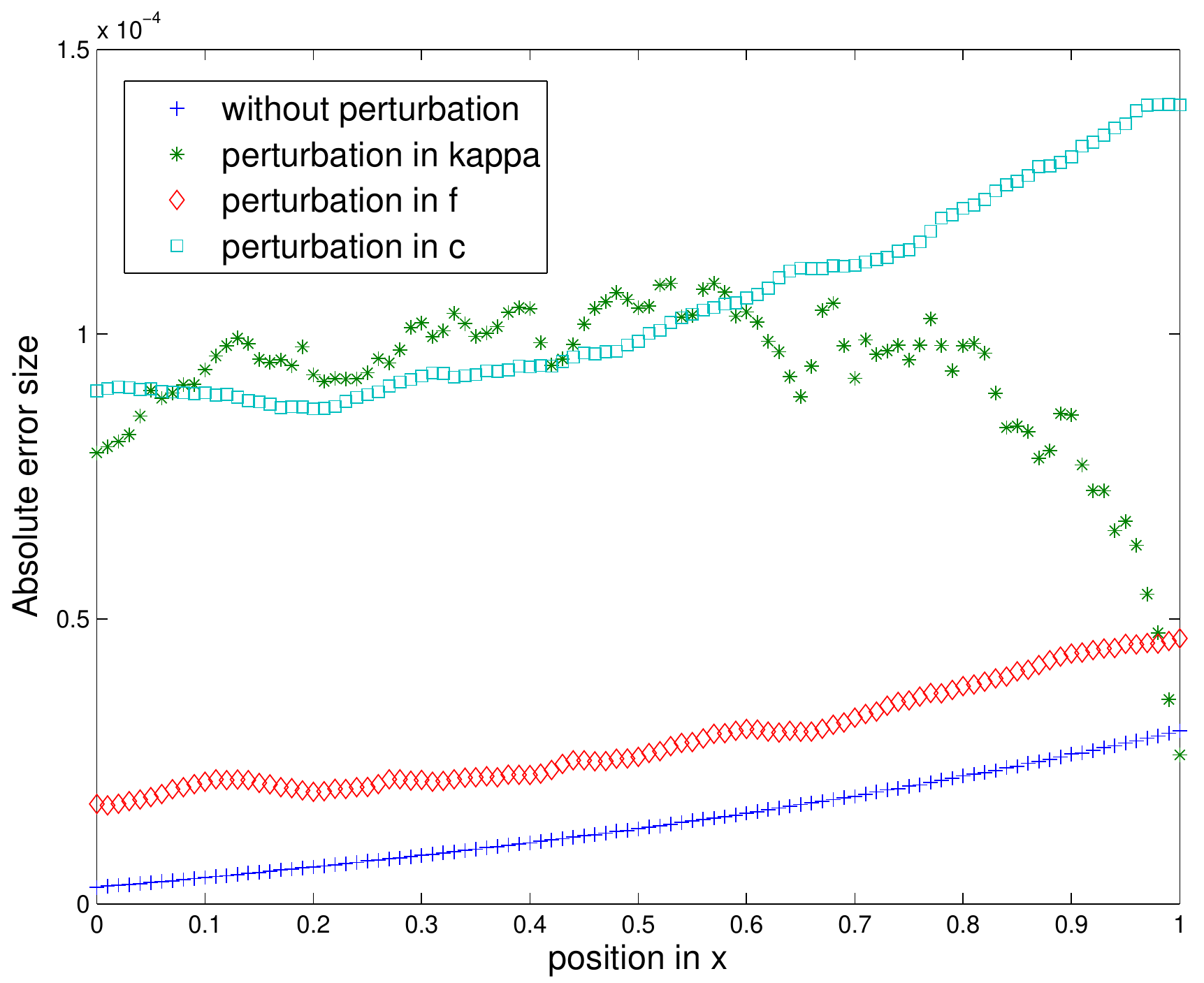}   %[scale=0.5]
         \caption{Absolute error for a fixed grid point index in $y$ direction.}\label{fig:perterr}
%       \end{minipage}

     \end{figure}

The results of this test are depicted in Figure \ref{fig:perterr} and quantified in Table \ref{perturbtest}. It seems that the perturbation of the right hand side $f$ has almost no influence on the numerical solution since the error of the perturbed equation is of the same magnitude as the error of the unperturbed setting. 

Perturbing $\kappa$ and $\tilde{c}$, however, does lead to an increase in the error of about one magnitude. Nevertheless, also in those cases the error is still below the discretization error and we need to keep in mind that the perturbation applied is several orders of magnitude larger than any oscillation introduced by round--off errors. 

%Then, the equation $- \Delta u + u = \tilde{f}$ is solved. The numerical solution should not noticeably differ from the analytical solution of the unperturbed equation. In this test, $100$ grid points are used in each direction.

%In figure \ref{noise}, the randomly perturbated right--hand side $\tilde{f}$ is plotted.
%The perturbation has no noticeable influence on the solution, as desired. The errors of the unperturbed and the perturbed solution compared with the analytical one are of the same size. Only in the $L^{\infty}$ norm, the error seems to be slightly larger in the perturbed case. Furthermore, the size of the error of the parallel solver is smaller than the discretisation error. This demonstrates the accuracy and the stability of the solver.

     \begin{table}
       \centering
       \begin{tabular}{|l|c|c|c|}
         \hline
         error & $L^1$ norm & $L^2$ norm & $L^{\infty}$ norm \\
         \hline \hline
         unperturbed, 1 core    & $1.4341 \times 10^{-5}$ & $1.6377 \times 10^{-5}$ & $3.0362 \times 10^{-5}$   \\
         unperturbed, 4 cores   & $1.4329 \times 10^{-5}$ & $1.6368 \times 10^{-5}$ & $3.0362 \times 10^{-5}$   \\
         $f$ perturbed, 1 core  & $2.8822 \times 10^{-5}$ & $2.9599 \times 10^{-5}$ & $4.9197 \times 10^{-5}$   \\
         $f$ perturbed, 4 cores & $4.5153 \times 10^{-5}$ & $4.6324 \times 10^{-5}$ & $6.5409 \times 10^{-5}$   \\
         \hline
         $\kappa$ perturbed, 1 core    & $10.946 \times 10^{-5}$ & $11.130 \times 10^{-5}$ & $23.476 \times 10^{-5}$   \\
         $c$ perturbed, 1 core          & $10.747 \times 10^{-5}$& $10.912 \times 10^{-5}$ & $15.839 \times 10^{-5}$   \\
         \hline
       \end{tabular}
       \caption{Poisson solver stability test: The numerical error in comparison with the analytical solution.}
       \label{perturbtest}
     \end{table}

%We repeated the test with perturbations of the coefficient functions $\kappa$ and $c$ of the same magnitude. The errors are about one magnitude larger %and vary significantly depending on the random perturbation, but are still small keeping the strength of the perturbation in mind. Since the error of %the parallelisation is about one magnitude smaller, we give only results for the serial case.

%     \begin{table}
%       \begin{tabular}{l|ccc}
%         error $\cdot 10^4$ & $L^1$ norm & $L^2$ norm & $L^{\infty}$ norm \\
%         \hline
%         $\kappa$ perturbed, 1 core  & 1.0946 & 1.1130 & 2.3476    \\
%         $\kappa$ perturbed, 4 cores & 0.8611 & 0.8720 & 1.4559    \\
%         $c$ perturbed, 1 core       & 1.0747 & 1.0912 & 1.5839    \\
%         $c$ perturbed, 4 cores      & 0.5153 & 0.5241 & 0.6816
%       \end{tabular}
%      \caption{Effect of perturbing $\kappa$ and $c$.}
%     \end{table}%-----------------------------------------------------------------------------------------------------------
\subsection{Time Stepping} \label{sec:timeintegration}

For time integration, the ANTARES framework provides the second and third order TVD Runge--Kutta schemes of \cite{tvdrk} as well as the second order three stage scheme of Kraaijevanger \cite{kraaijevanger1991}. N. Kwatra et al. suggested 
in \cite{kwatra} two variations of the temporal integration with Runge--Kutta methods. The first is to perform Runge--Kutta on just $F_{\mathrm{ad}}(U)$ and 
$F_{\mathrm{vis}}(U)$ with one final implicit integration of $F_{\mathrm{nonad}}(U)$. For the second variation they suggest to integrate 
$F_{\mathrm{ad}}(U)$, $F_{\mathrm{vis}}(U)$ and $F_{\mathrm{nonad}}(U)$ for each Runge--Kutta stage. In \cite{kwatra} a better performance is reported 
employing the second variation. Since this coincides with our observations, we use the second variation in all our simulations. 

\vspace{1em}

In \cite{kwatra} a Courant--Friedrich--Levy condition based on an estimate of the maximum value of $\left|u\right|$ and the pressure gradient $\nabla P$ 
throughout the next timestep is derived,

\begin{equation}
\Delta t \left(\frac{\left|u\right|_{\mathrm{max}} + \Delta t \frac{\left|\nabla P\right|}{\rho}}{\Delta x}\right) \leq 1 .
\end{equation}

The term corresponding to the pressure gradient can be neglected in the limit where $\Delta x \rightarrow 0$, since its contributions are of lower order (see \cite{strikwerda} for a discussion of the treatment of lower order terms in stability analysis). In that case, we arrive at a timestep restriction exclusively due to the velocity of the flow,

\begin{equation}
\tau_{\mathrm{fluid}} = CFL_{\mathrm{adv}} \frac{\Delta x }{\left|u\right|_{\mathrm{max}}}
\end{equation}

Further restrictions on the timestep $\Delta t$ are imposed by heat diffusion $\tau_{\mathrm{T}}$ and the viscosity $\tau_{\mathrm{visc}}$. In case of a two 
component flow, the diffusion of the second component also leads to a timestep restriction, $\tau_{\mathrm{c}}$. 

We define our timestep as 

\begin{equation}
\Delta t = \mathrm{min}\left\{\tau_{\mathrm{T}},\tau_{\mathrm{visc}}, \tau_{\mathrm{fluid}}(,\tau_{\mathrm{c}})\right\} 
\end{equation}

where

\begin{equation*}
\tau_{\mathrm{c}} = CFL_{\mathrm{diff}} \cdot \frac{\Delta x^2}{\kappa_{\mathrm{c}}}, \qquad
\tau_{\mathrm{T}} = CFL_{\mathrm{diff}} \cdot \frac{\Delta x^2}{\kappa_{\mathrm{T}}}, \qquad
\tau_{\mathrm{visc}} = CFL_{\mathrm{diff}} \cdot\frac{\Delta x^2}{\nu}. \label{eq:taudiff}
\end{equation*}

$CFL_{\mathrm{diff}}$ and $CFL_{\mathrm{adv}}$ denote Courant numbers not necessarily equal. 

The maximum Courant numbers $CFL_{\mathrm{adv\; max}}$ and $CFL_{\mathrm{diff\; max}}$ may be determined by (linear) von Neumann stability analysis 
(see, for example \cite{strikwerda}). However, in case of $CFL_{\mathrm{adv \; max}}$, the nonlinear WENO scheme must be linearized and the maximum 
Courant number obtained by linear stability analysis provides just an estimate. In fact, the maximum 
Courant number of WENO5 combined with the third order TVD Runge--Kutta method is predicted to be $1.43$, but simple tests indicate that this solver 
becomes unstable at $CFL_{\mathrm{adv}} = 1.0$ (see \cite{MotamedRuuth2011} and \cite{WangSpiteri2007}). Thus, in practice we make sure to choose a 
Courant number smaller than the one theoretically possible.    

\vspace{1em}

The source term on the right hand side of equation (\ref{eq:navstoonecom}), which represents buoyancy forces acting on the flow, also poses a restriction on the timestep, but similarly to the restriction caused by the pressure gradient above its contributions are of lower order and hence typically only provide an accuracy but not a stability restriction.

%-----------------------------------------------------------------------------------------------------------
\section{Simulations in the Astrophysical Regime}
\subsection{Physical Setting}
To set up a model of stellar convection, we follow
the ansatz of \cite{muthsametal1999} and \cite{zaussinger} by considering a hydrostatically layered fluid column which is unstable against convection.  In terms of the Schwarzschild criterion of convective instability \cite{coxgiulies} with x pointing along the direction of gravity, this reads
\begin{equation}
\frac{\partial T}{\partial x} > \left(\frac{\partial T}{\partial x}\right)_{\mathrm{ad}} .
\end{equation} 

From the outset we assume the gas to be ideal gas with a $\gamma$--law where $\gamma =5/3$. The volume expansion coefficient is given by $\alpha = 1/(\gamma-1)$ and the specific heat at constant pressure is $c_{\mathrm{P}} = R_{\mathrm{gas}} (1+\alpha)$. 

These data suffice to determine the adiabatic temperature gradient
\begin{equation}
\left(\frac{\partial T}{\partial x}\right)_{\mathrm{ad}} = \frac{g}{c_{\mathrm{P}}} 
\end{equation}
once we have specified the downward pointing constant gravity $g$. 

We fix the temperature at the top and define a function $b(x)$ to be the ratio of the actual temperature gradient to the adiabatic one:
\begin{equation}
\frac{\partial T}{\partial x} = b(x) \left(\frac{\partial T}{\partial x}\right)_{\mathrm{ad}} . \label{bfac}
\end{equation}

We set $b(x) = a$ with $a>1$ and  integrate equation (\ref{bfac}) straightforwardly. This way the temperature -- and therefore the energy -- 
is specified for the entire domain. 

We determine the density at the top and eliminate the pressure $P$ in the equation of hydrostatic equilibrium. Simple integration yields the density for the entire domain. The pressure $P$ is obtained from the equation of state (\ref{eq:idealEOS}). 

\vspace{1em}

To complete our model we need to specify the viscosity coefficient $\eta$ and the radiative conductivity $\kappa$. This is achieved by prescribing the 
Rayleigh number $Ra$ and the Prandtl number $Pr = c_{\mathrm{P}} \eta / K$. The former quantities arise in the definition of the starting model only, since in our 
model we assume $K(x,t)$ and $\eta(x,t)$ to be constant. 
This setup simplifies studies of the basic physics while it is still useful for insight into in the astrophysically relevant case. 

\vspace{1em}

To start dynamics away from equilibrium we apply a random initial perturbation. 

\subsubsection{Starting model for the semiconvection problem}

Double--diffusive convection is a phenomenon encountered in chemically inhomogeneous flows where a significant mean molecular weight gradient is present.  

Semiconvection is the most important special case of double--diffusive convection in astrophysics. Models of stellar structure and evolution predict 
settings where the heavier product of nuclear fusion provides stability to a zone which otherwise would be unstable to convective overturning, because 
temperature sufficiently rapidly decreases against the direction of gravity. Such a zone would become convective, if its composition were mixed, and the 
question whether such a zone should be treated as if it were mixed or not has become known as the \textit{semiconvection problem} \cite{zaussispruit}. 

\vspace{1em}

To model a chemically inhomogeneous fluid we need to add an equation describing the dynamics of the second species to our set of equations. 
The corresponding partial density equation has been introduced in Section \ref{sec:numericalmethod}. The starting model for a semiconvective simulation 
has been developed by F. Zaussinger \cite{zaussinger} and is set on top of the starting model for stellar convection by expanding it with the 
specification of the Lewis Number $Le = c_{\mathrm{P}} \rho \kappa_{\mathrm{c}}/K$, which relates the radiative conductivity $K$ to the diffusion 
coefficient $\kappa_{\mathrm{c}}$ as well as with the stability parameter $R_{\mathrm{\rho}}$. The stability parameter is defined as the ratio 
of the gradient of mean molecular weight to the superadiabatic temperature gradient (which itself is given by b(x)-1 from equation (\ref{bfac}). Further details are found in \cite{zaussinger}. 

\subsection{Simulation Setting}

Due to the compressible nature of the flows we model, we specify the vertical extent of the simulation domain in multiples of the pressure scale height $H_{\mathrm{P}}=P/(\rho g)$. For the simulations presented the domain always covers 1 $H_{\mathrm{P}}$. 

\vspace{1em}

To determine the resolution required to resolve even the smallest structures in our simulations -- namely, to perform a \textit{direct numerical simulation} -- 
we resort to mixing length theory. In \cite{zaussinger}, a correlation between the Rayleigh number and the diffusion coefficients is derived and a 
balance between advection by the interior flow and the diffusion across an interface is used to estimate the boundary layer thickness $\delta_{\mathrm{T}}$ as

\begin{equation}
\delta_{\mathrm{T}} = Ra^{*\;-\frac{1}{4}} \; H , \label{eq:Tboundary}
\end{equation}

where H denotes the (physical) height of the domain and $Ra^* = Ra \cdot Pr$ is the modified Rayleigh number.
If H denotes the number of grid points along the vertical direction, then $\delta_{\mathrm{T}}$ is the number of grid points resolving the thermal boundary layer. 

\vspace{0.5em}

Similarly, in case of double--diffusive convection, the boundary layer thickness of the second species $\delta_{\mathrm{c}}$ follows from the definition of the Lewis Number and scales with the thermal boundary layer thickness,

\begin{equation}
\delta_{\mathrm{c}} = \sqrt{Le} \; \delta_{\mathrm{T}} . \label{eq:Heboundary}
\end{equation}

The same conclusion can be drawn for the viscous boundary layer, 
\begin{equation}
\delta_{\mathrm{visc}} = \sqrt{Pr} \; \delta_{\mathrm{T}} , 
\end{equation}
using here the definition of the Prandtl number $Pr$. 

\vspace{1em}

In our simulations we seek to resolve the boundary layers with a minimum of 5 grid points. 

\vspace{1em}

The system of equations is discretized on a rectangular, equidistantly spaced grid. 
Boundary conditions are based on the assumption that all quantities are periodic in the horizontal direction. 
Moreover, for the hydrodynamical equations, we employ ``closed'' (Dirichlet) boundary conditions in the vertical direction. 

\vspace{1em}

The simulation time is measured in units of \textit{sound crossing times (scrt)}. One sound crossing time is defined as the time taking an 
acoustic wave to propagate from the bottom to the top of the simulation box.

%-----------------------------------------------------------------------------------------------------------
\subsection{A Numerical Instability} \label{sec:numericalresults1}
After ascertaining the basic properties of the numerical method as outlined in \cite{kwatra}, we extended the method to two-component viscous flow and turned to simulations of stellar convection and 
stellar semiconvection. Although for some parameter sets the extended method performed well and exhibited the higher timestep expected from it, for other 
parameter sets spurious oscillations appeared in the numerical solution. 

\begin{figure}
\centering
\includegraphics[width=9cm, height=6cm]{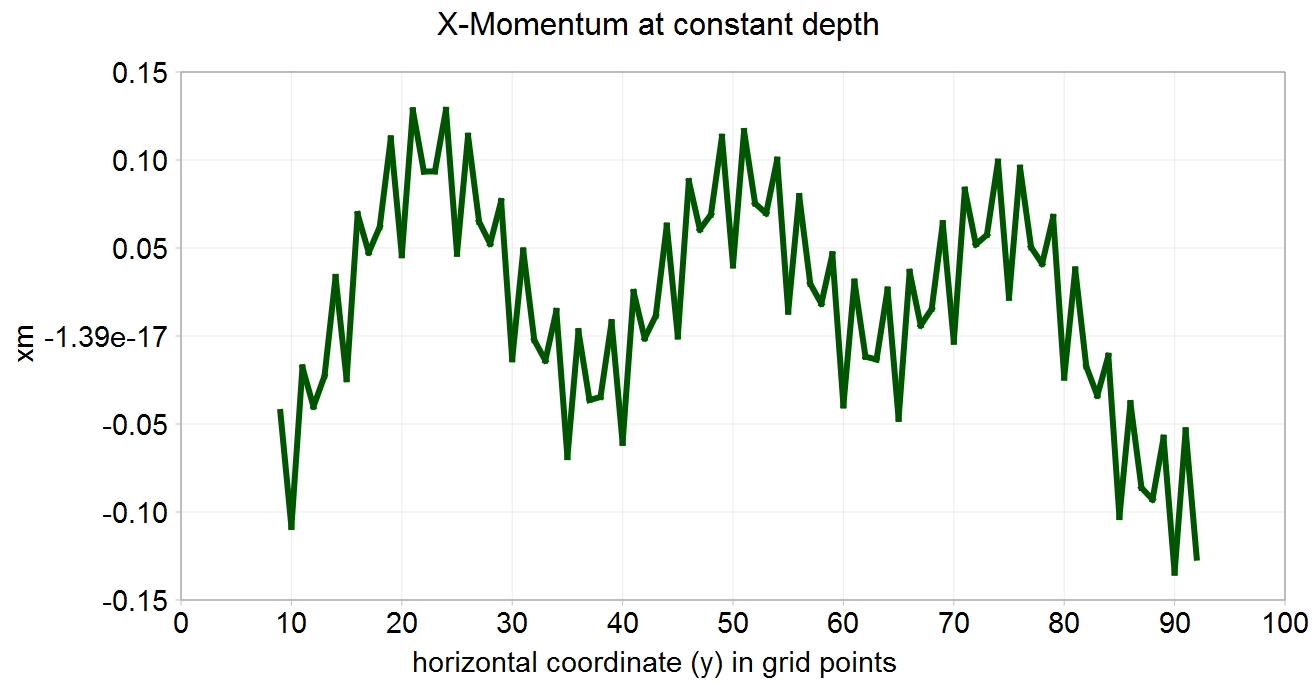}

\caption{Simulation of double--diffusive convection. The upper picture shows a horizontal slice of the momentum in x--direction after 47 scrt. Already, two--point instabilities have appeared.}
\label{fig:KSCNOBFL}
\end{figure}

In Figure \ref{fig:KSCNOBFL} we show a snapshot of a simulation of double--diffusive convection. The simulation parameters are 
$Pr=1.0$, $Le=1.0$, $R_{\mathrm{\rho}}=1.1$ and $Ra^* = 160 000$. We have a spatial resolution of 100 $\times$ 100 grid points and choose a 
CFL Number of $CFL_{\mathrm{adv}}=CFL_{\mathrm{diff}} = 0.15$. After a certain time, two--point instabilities appear in the numerical solution, 
propagate over the entire domain and render the simulation useless. 

The appearance of the instabilities was quite surprising since the explicit counterpart of the method did not show any similar instability effects. 
We therefore strived to enhance the long--term stability of our solver. 

%-----------------------------------------------------------------------------------------------------------

\section{Enhancing Stability: Dissipative Spatial Discretization} \label{sec:dissipativestencil}
In the ANTARES framework, the second--order terms have been evaluated via a finite difference scheme of fourth order \cite{antares,SDIRKreport},

\begin{eqnarray*}
\lefteqn{(U_i^n)_{xx} =} \\
& & \frac{-U^n_{i+4} -18 U^n_{i+3}+ 80 U^n_{i+2} + 18 U^n_{i+1} - 162 U^n_{i} + 18 U^n_{i-1} + 80 U^n_{i-2} - 18 U^n_{i-3} + U^n_{i-4}}{192 h^2} \\
& & +{ }O(h^4) \label{eq:nondissipativestencil} 
\end{eqnarray*}

This stencil satisfies the properties of flux conservation, easy generalization to more spatial dimensions through directional splitting and easy 
implementation into conservative numerical schemes for the hyperbolic part of the hydrodynamical equations. However, the following analysis shows that 
this stencil lacks a crucial property: strict dissipativity. 

\subsection{Dissipativity Analysis}

A strictly dissipative finite difference scheme efficiently damps high--frequency oscillations in the numerical solution.
To determine whether our scheme is strictly dissipative or not, we resort to dissipativity analysis (see \cite{strikwerda} for the general idea). 

\vspace{1em}

The crucial quantity studied in this context is the \textit{amplification factor $g=g(\theta)$}. The propagation of the numerical solution 
by one time step corresponds to the multiplication of the Fourier transform of the numerical solution by this factor $g$. Thus, the magnitude of 
the amplification factor equals the factor by which the amplitude of each frequency in the solution is increased or decreased during each time step. 
Hence, the finite difference scheme is strictly dissipative if $ \left|g(\theta)\right| < 1 $ for all $\theta \in [-\pi,\pi]$ and $g(0)=1$. In general, $\theta$ 
depends on the timestep $\Delta t$ and, therefore, $g(\theta)$ depends on the Runge--Kutta method used for time integration. 

\vspace{1em}

The detailed analysis is given in \cite{SDIRKreport}, so we will limit ourselves to the results. It turns out that for both the second and third order 
TVD Runge--Kutta methods, $g(\theta) = 1$ for $\theta \in \left\{-\pi,0,\pi\right\}$ while $\left|g(\theta)\right| < 1$ for sufficiently small $\Delta t$ for all other $\theta \in [-\pi,\pi]$. Thus, both schemes reduce the magnitude of most frequencies 
(given an accordingly chosen timestep) but not the highest frequencies appearing on the grid scale. This is essentially the property we lack in our simulations. 

Therefore, we here derive a finite difference scheme which is dissipative, of fourth order and consistent with the ENO spatial discretization employed for the advective terms of the Navier--Stokes equations. 
 
%-----------------------------------------------------------------------------------------------------------
\subsection{Derivation of the finite difference discretization from the WENO approach}

For the convenience of the reader, we repeat the derivation suggested in \cite{SDIRKreport}. 

In the WENO--approach \cite{eno1} the $U_i$ are assumed to be cell averages of a function $v(x)$,
\begin{equation}
U_i = \frac{1}{h} \int\limits_{x_{i-\frac{1}{2}}}^{x_{i+\frac{1}{2}}}{v(\zeta) \, d\zeta} \label{eq:fv-approach}
\end{equation}
Approximating $v(x)$ with a polynomial function $p(x)$ to specified order, $p(x)=v(x)+O(h^5)$, one obtains
\begin{equation}
(U_i)_x = \frac{1}{h} (p(x_{i+\frac{1}{2}})-p(x_{i-\frac{1}{2}})) + O(h^5)
\end{equation}
as long as the functions are smooth enough to give an extra $h$ in the 
difference to cancel that one in the denominator. 
Clearly, using $p(x_{i+1/2})$ as approximation to $U_{i+1/2}$ has a lower order of accuracy, the
higher order comes from perfect cancellation when building the difference.
Now using $p'(x_{i\pm1/2})$ for building the 2nd derivative one obtains
\begin{equation}
(U_i)_{xx} = \frac{p'(x_{i+\frac{1}{2}})-p'(x_{i-\frac{1}{2}})} {h}.
\end{equation}
%ie. setting $(\hat{f}_{i+1/2})^n = p'(x_{i+1/2})$ from the previous section. 
Centered 4--point stencil approximation leads to 
\begin{equation}
        p_{i+\frac{1}{2}}    = \frac{U_{i-1} - 15\,U_{i} + 15\,U_{i+1} - U_{i+2}}{12\,h} + O(h^2), 
\end{equation}
and
\begin{equation}
        p_{i-\frac{1}{2}}    = \frac{U_{i-2} - 15\,U_{i-1} + 15\,U_{i} - U_{i+1}}{12\,h} + O(h^2), \label{eq:fv-approach2} 
\end{equation}
where the order of accuracy is measured with respect to $(U_{i\pm 1/2})_x$. 
The calculation was done using a Mathematica 4.1 script.
The second derivative is then given by
\begin{equation}
(U_i^n)_{xx} = \frac{-U_{i-2} + 16\,U_{i-1} - 30\,U_{i} + 16\,U_{i+1} - U_{i+2}}{12\,h^2} + O(h^4) \label{eq:dissipativestencil4} 
\end{equation}
which corresponds to the standard centered 5--point stencil approximation.

\vspace{1em}

The corresponding dissipativity analysis in \cite{SDIRKreport} shows that this spatial discretization is dissipative in combination with the Runge--Kutta methods from \cite{tvdrk} mentioned in Section \ref{sec:timeintegration}(the method of \cite{kraaijevanger1991} mentioned there is analyzed in the appendix of \cite{imex2012}). However, this property comes at the cost of an additional restriction to the stability constraint. Monotonic decay of the numerical solution is ensured, if the diffusive Courant Number in (\ref{eq:taudiff}) satisfies
 
\begin{equation}
CFL_{\mathrm{diff \; 2 \; max}} = \frac{3}{4} CFL_{\mathrm{diff \; max}}
\end{equation} 

\vspace{1em}

The reader will notice that the dissipativity analysis has been performed assuming (\ref{eq:dissipativestencil4}) to be a finite difference scheme, 
whereas (\ref{eq:dissipativestencil4}) was derived in a finite--volume framework (e.g. equation (\ref{eq:fv-approach})). The justification for mixing 
both approaches is given by Shu in \cite{Shu2009}. He shows that for one--dimensional conservation laws the finite--volume and the finite--difference WENO 
reconstruction procedure is essentially the same. The only difference is found in the input--output pair: for a finite volume scheme, the input is the set 
of cell averages and the output is the set of reconstructed values of the solution at the cell interfaces; for a finite--difference scheme, the input is the
set of point values of the physical flux and the output is the set of numerical fluxes at the cell interfaces. The transition to several dimensions is obtained by 
dimensional splitting as common in finite difference methods. Note that the important part of the derivation given here is actually the intermediate result (\ref{eq:fv-approach})--(\ref{eq:fv-approach2}): this is the way the stencil is implemented in ANTARES and it ensures the property of flux conservation and easy implementation into the WENO-framework when combined with dimensional splitting. 

\section{Numerical results and further tests}

In all the following simulations we use the second order TVD Runge-Kutta method of \cite{tvdrk} for time integration. We demonstrate the potential of the implicit time integration of the pressure terms through the approach of \cite{kwatra} and with the extensions derived in the previous sections for a number of problems including idealized cases, benchmark tests and astrophysical applications. We also show that the method performs extremely well on highly parallelized supercomputing platforms. 

\subsection{Speed-up of simulations of convection and semiconvection}

\begin{figure}
\centering
\includegraphics[width=9cm, height=6cm]{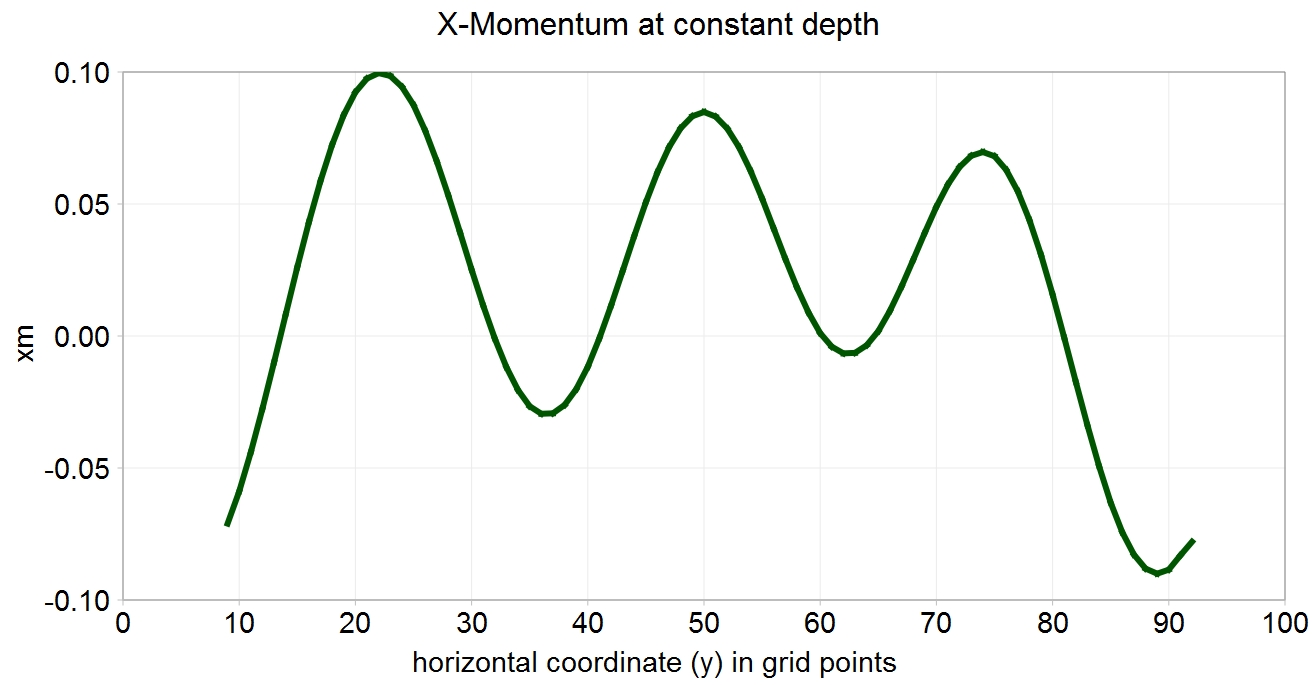}
\caption{Rerun of the simulation described in Section \ref{sec:numericalresults1} with strictly dissipative spatial discretization. The picture shows a horizontal slice of x--momentum after 47 scrt. There is no indication of spurious oscillations.}
\label{fig:KSCBFL}
\end{figure}

We first rerun the simulation of double--diffusive convection shown in Figure \ref{fig:KSCBFL} employing the dissipative finite difference scheme derived in the previous section. Although the new stencil entails a more severe timestep restriction, we choose the Courant number such that in both simulations the same timestep is used. We see in Fig. \ref{fig:KSCBFL} that due to the property of strict dissipativity our improved solver efficiently damps the high frequency oscillations on the grid scale and leads to a smooth and stable solution. 

%It was surprising to find the origin of the mentioned instabilities in the non-dissipative spatial discretization scheme, since the fully explicit solver 
%which consists of WENO-scheme for evaluation of the hyperbolic part of the NSE and the non-dissipative finite difference scheme (\ref{eq:nondissipativestencil}) 
%did not exhibit similar features for the simulation mentioned in Section \ref{sec:numericalresults1}. However, the fully explicit solver is 
%limited by the timestep induced by the sound speed, which is significantly smaller than the timestep permitted by the method of \cite{kwatra}. %Furthermore,
%we also note that some amount of numerical dissipation on the grid scale is provided by the WENO scheme.
%One property of the Poisson-like equation is that information is propagated infinitely fast, such that any instability appearing in 
%the pressure evolution equation travels instantaneously through the entire domain. Both facts, the large timesteps and the error propagation through 
%(\ref{eq:GeneralizedPoisson}) may contribute to the fast propagation and build-up of the instabilities appearing on the grid scale. 

As we have already mentioned, the explicit counterpart of the fractional step method, namely, employing the WENO scheme for the entire hyperbolic part and discretizing the viscous terms via the dissipative scheme (\ref{eq:nondissipativestencil}) has rendered stable results. Similarly, the Boussinesq equations discretized with the same techniques as the explicit solver in \cite{zaussinger} yielded stable results even at much higher timesteps. For in the present fractional step method we suspect that some instability is introduced by the discretization of equation (\ref{eq:pressev1}), where the advective term and the source term are discretized using an explicit Euler--forward timestep whereas the diffusive pressure is defined implicitly at the new timestep.

However, we rule out the solver for the Poisson--like equation as a sole initiator of the instabilities, partly, because the same solver is successfully employed solving the Boussinesq equations in \cite{zaussinger} and partly due to the satisfyingly stable results the solver renders even, if slight oscillations are present in the coefficient functions (see Section (\ref{sec:StabilityPoisson})). Nevertheless, one property of the generalized Poisson equation is that information is propagated infinitely fast such that any instabilities appearing in the numerical solution travels instantaneously through the entire domain. This might favor the build--up of the observed oscillations. 

We conclude that the fractional step method may be stabilized by employing a strictly dissipative discretization scheme.

\vspace{1em}

Having successfully enhanced the longterm stability of the method, we performed a series of simulations of stellar convection and double--diffusive convection. 
Such a simulation is expected to evolve the following way: after the initial vertical oscillations\footnote{\label{foot:1} These oscillations 
arise because due to rounding errors and due to the different truncation errors of the integration of the initial condition and the spatial discretization of 
the dynamical equations in the setup of the starting model the simulation is not in perfect hydrostatic equilibrium.} are damped out, the velocity field slowly starts 
building up. Large scale gravity waves form and break, causing the fluid to mix locally. The turbulent mixing process spreads over the whole domain and 
after a while, quasi--stationary convective rolls form.

In case of the double--diffusive convection scenario, the main difference to the chemically homogeneous convective scenario is that the stable mean 
molecular weight gradient delays the build--up of the velocity field, leading to a considerable prolongation of the diffusive phase. By the term ``diffusive 
phase'' we refer to the first part of the simulation, where the fastest timescale is set by diffusion processes. Eventually, the fluid velocity 
increases such as to pose an even greater restriction on the timestep than diffusion. Once this happens, the simulation has reached the ``advective phase''.

\vspace{1em} 

\begin{figure}[ht!]
\centering
\includegraphics[scale=0.3, angle=270]{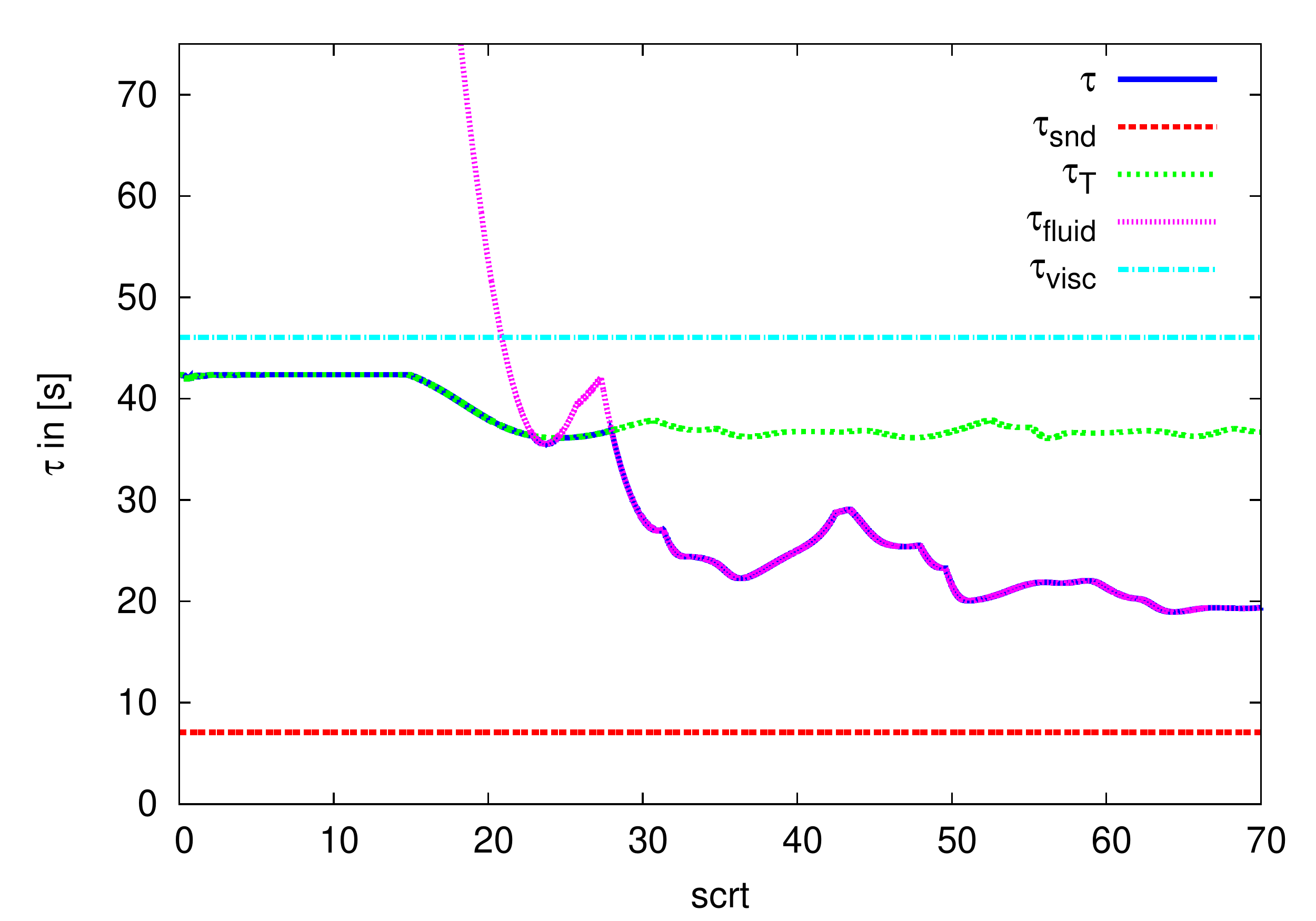}
\includegraphics[scale=0.3, angle=270]{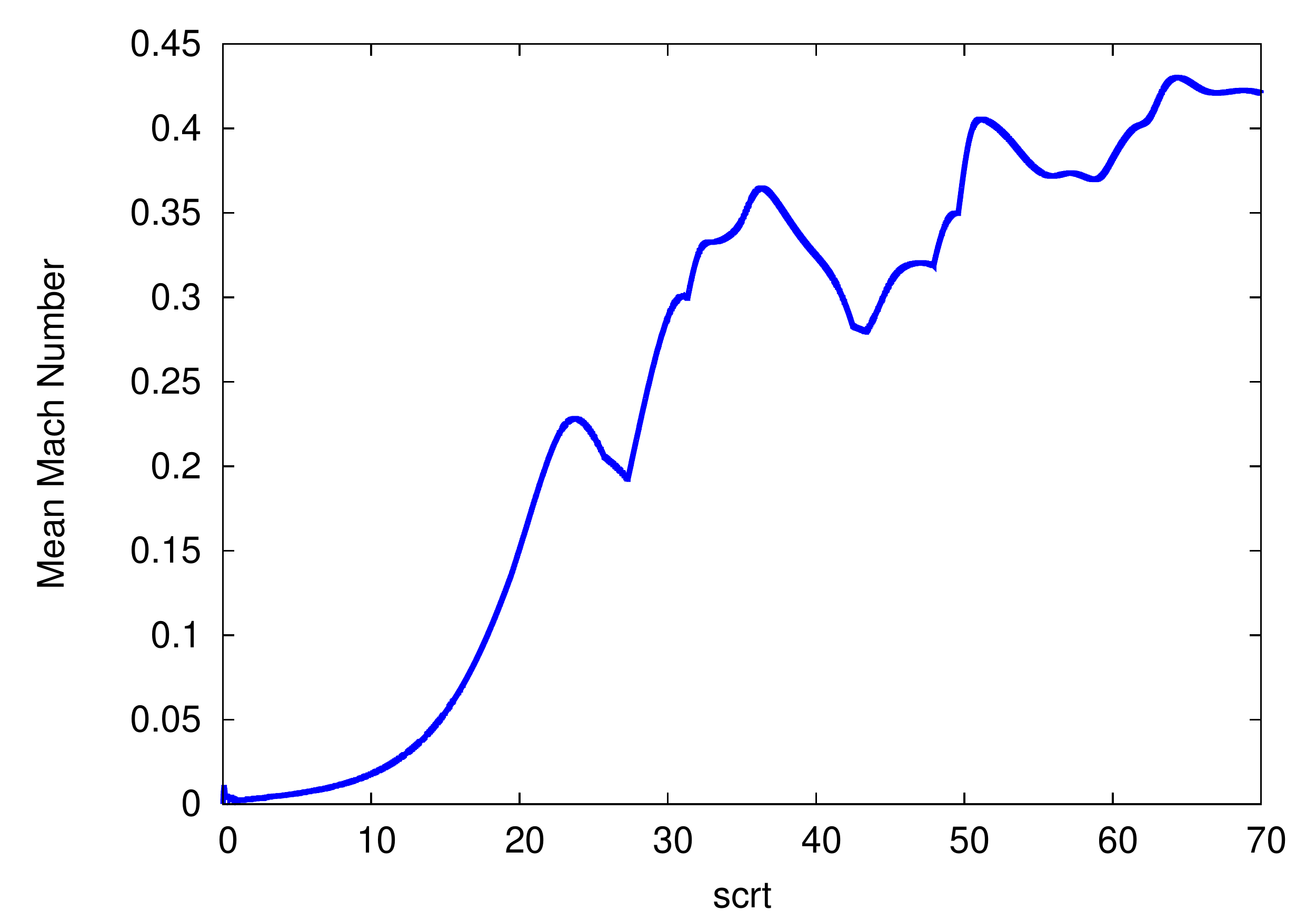}
\caption{Simulation of stellar convection with parameters $Pr=1.0$ and $Ra^* = 160 000$. The spatial resolution is 150 $\times$ 150 grid points and we use a 
Courant number $CFL_{\mathrm{adv}} = CFL_{\mathrm{diff}} = 0.2$. Simulation time is 70 scrt.} 
\label{fig:Conv}
\end{figure}

Figure \ref{fig:Conv} shows the timestep evolution in a simulation of stellar convection. The maximum timestep advantage compared to the explicit scheme
is achieved in the diffusive phase of the simulation, which lasts in this case approximately 30 scrt. The build--up of the velocity field is best 
followed by looking at the mean Mach number. In the range of 10--20 scrt the velocity steadily increases, until the first gravity waves break at 
approximately 22 scrt. The mixing process slows the growth of the velocity until it reaches a steady state of convective motions (at approximately 55 scrt). 
At a certain point, the fluid velocity is high enough to pose a greater restriction on the timestep than diffusion. Nevertheless, even in the final state with quasi--stationary
averaged properties the timestep employed is still twice the sound--speed based timestep. 

\begin{figure}[ht!]
\centering
\includegraphics[scale=0.3, angle=270]{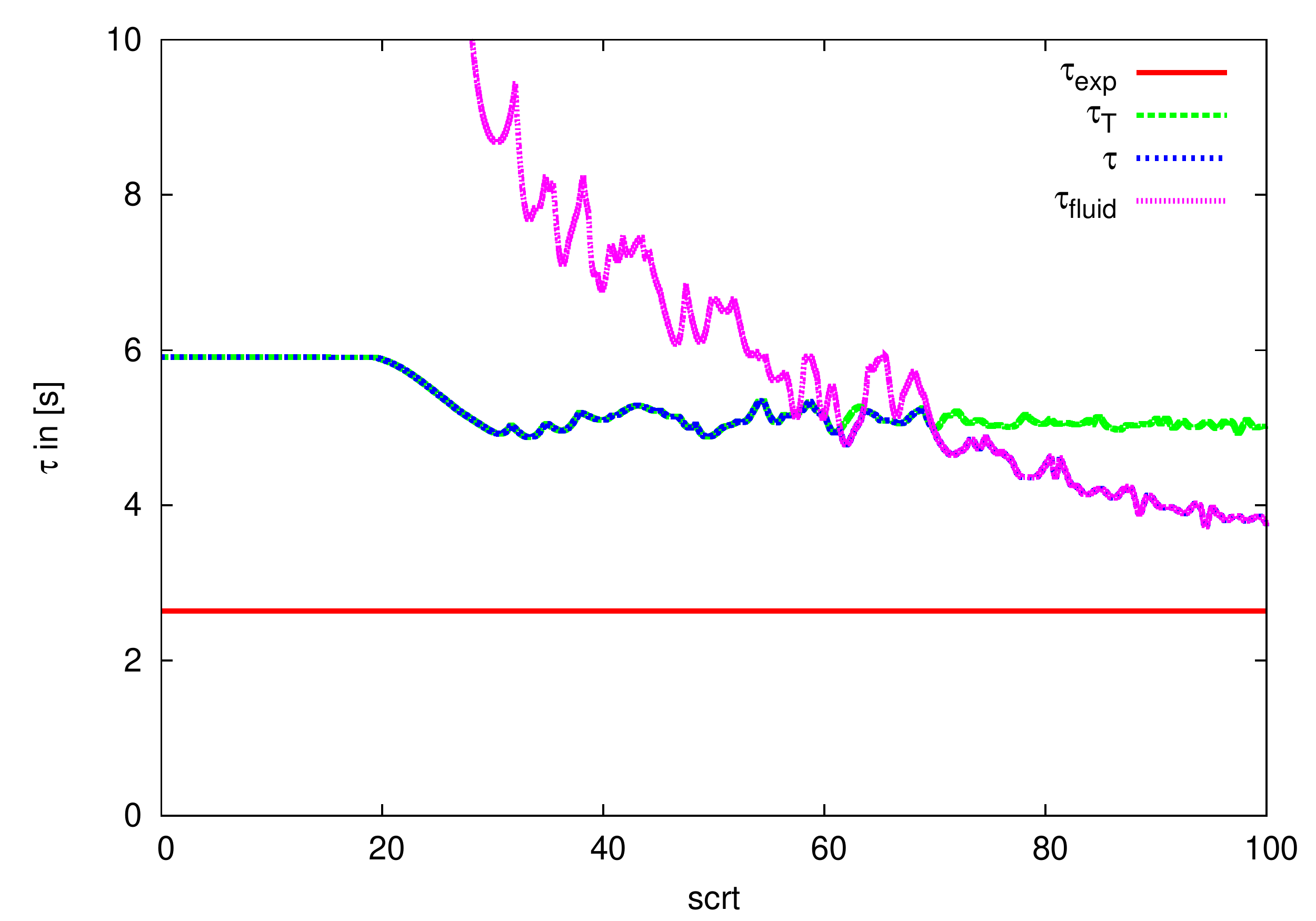}
\includegraphics[scale=0.3, angle=270]{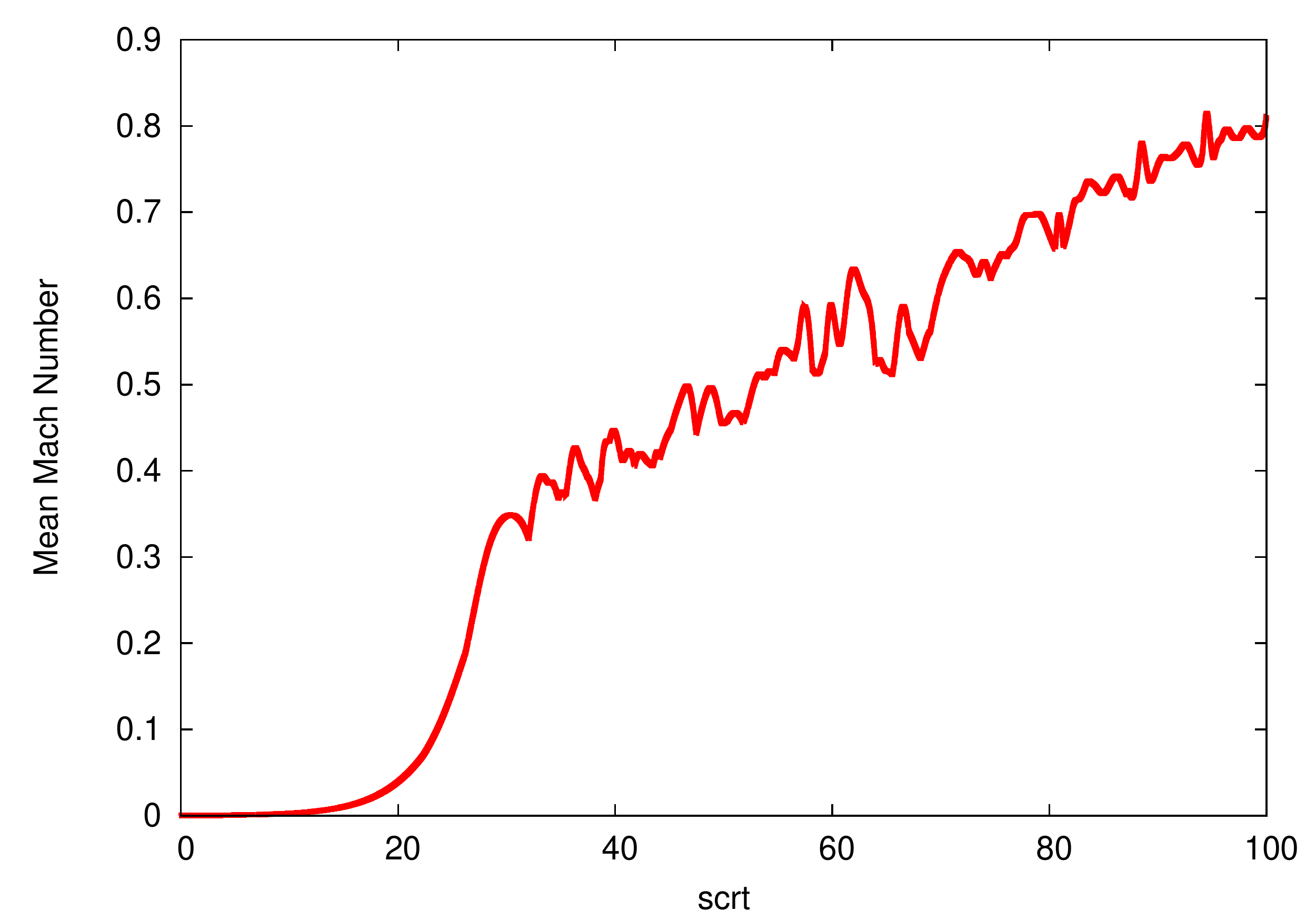}
\caption{Simulation of stellar convection with parameters $Pr=0.1$ and $Ra^* = 160 000$. The spatial resolution is 400 $\times$ 400 grid points and we use a 
Courant number $CFL_{\mathrm{adv}} = CFL_{\mathrm{diff}} = 0.2$. Simulation time is 100 scrt.} 
\label{fig:Conv2}
\end{figure}

Altering the parameters of the setting of stellar convection, however, leads to a simulation where much higher fluid velocities are encountered. Although the efficiency gain in the timestep is limited to the initial diffusive phase, we give this example to show that unlike methods especially tailored to the low Mach number regime, the fractional step method is capable of passing smoothly to regimes where Mach numbers of almost 1 are encountered. The evolution of the mean Mach number as well as the timesteps are shown in Figure \ref{fig:Conv2}.  

\vspace{1em}

\begin{figure}[ht!]
\centering
\includegraphics[scale=0.3, angle=270]{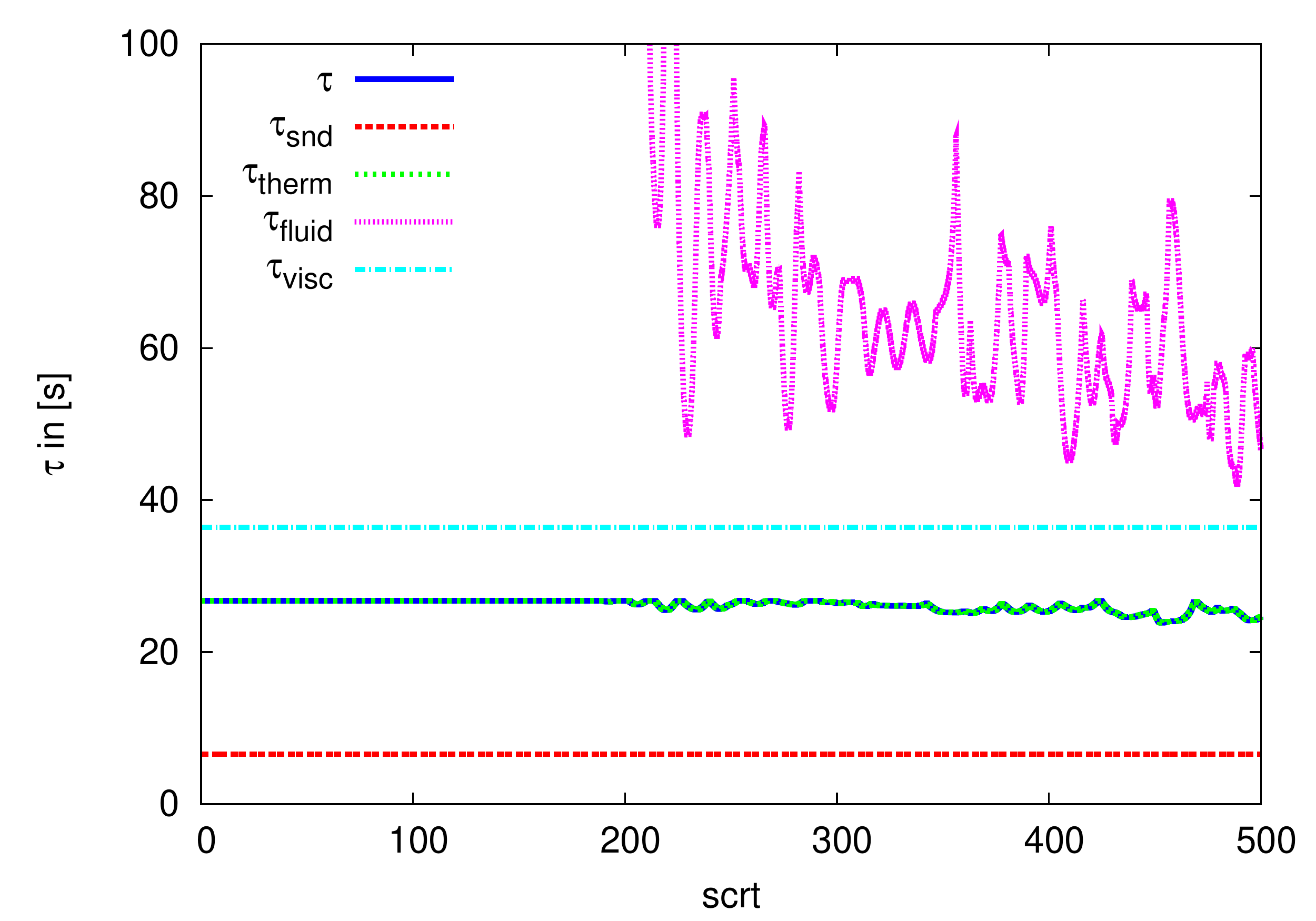}
\includegraphics[scale=0.3, angle=270]{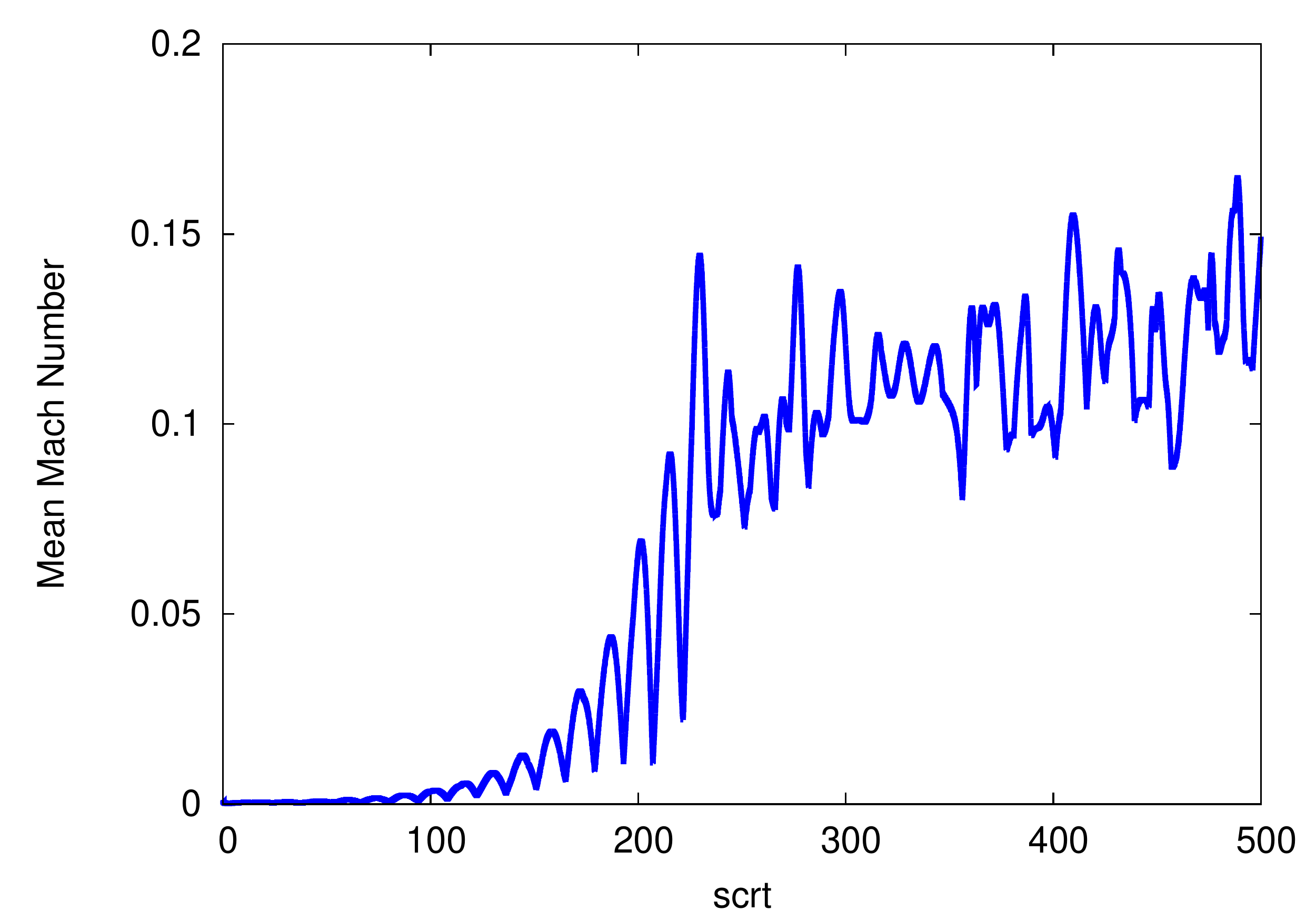}
\caption{Simulation of double--diffusive convection with parameters $Pr=1.0$, $Ra^* = 160 000$, $Le = 0.5$ and $R_{\rho} = 1.1$. The spatial resolution is 
150 $\times$ 150 grid points and we use a Courant number $CFL_{\mathrm{adv}} = CFL_{\mathrm{diff}} = 0.2$.}
\label{fig:SC} 
\end{figure}

Figure \ref{fig:SC} depicts the timestep evolution in a simulation of semiconvection. Here we encounter a real low Mach number regime since the Mach number 
never exceeds 0.2 and we see that the relevant timestep restriction stems from diffusive processes which leads to a timestep nearly 5 times higher 
than $\tau_{\mathrm{snd}}$. Table \ref{timesteps} lists the mean and maximum timesteps achieved in the simulations.

\begin{table}[ht!]
 \centering
 \begin{tabular}{|c||l|l|l|l|l|}
\hline
  setting & $\tau_{\mathrm{max}}$ & $\tau_{\mathrm{mn}}$ & $\tau_{\mathrm{exp}}$ & $\tau_{\mathrm{max}}/\tau_{\mathrm{exp}}$ & $\tau_{\mathrm{mn}}/\tau_{\mathrm{exp}}$  \\\hline
  \hline
  Convection, Pr=1.0 & 42.36 s & 32.03 s & 7.07 s & 6.0 & 4.53 \\\hline
  Convection, Pr=0.1 & 5.91 s & 3.73 s& 2.64 & 2.24 s &  1.41 \\\hline
  Semiconvection & 26.73 s& 26.5 s & 6.57 s & 4.07  & 4.03 \\
  \hline
 \end{tabular}
\caption{Maximum and mean timesteps ($\tau_{max}$ resp. $\tau_{mn}$)  achieved in the simulation of convection and double diffusive convection. To highlight the gain in performance, we give the ratio of the achieved timesteps and the sound speed based timestep $\tau_{\mathrm{exp}}$. }
 \label{timesteps}
\end{table}

\subsection{Convergence Test}
To demonstrate the accuracy of our solver, we performed a grid refinement study of a double--diffusive single layer simulation with parameters $Pr=0.1$, $Le=0.1$, $R_{\rho}=1.1$ and $Ra^*=1.6\cdot 10^6$. We successively increased the resolution, starting at a simulation with 200$\times$200 grid points and arriving at a maximum resolution of 800$\times$800 grid points. In view of our criteria (\ref{eq:Tboundary}) and (\ref{eq:Heboundary}) for resolving also the smallest structures in our simulations, the coarsest grid resolves the Helium boundary layer with about 3 grid points (which is quite low) whereas on the finest grid, the diffusive layer is resolved with 12 grid points. 

We compare the error between sucessive grid resolutions, i.e. we compare the error between the simulations with 200$\times$200 grid points and 400$\times$400 grid points to the error between the simulations with 400$\times$400 grid points and 800$\times$800 grid points. As expected, the error between successively refined grids decreases. This strongly indicates that our method is convergent. 
%We take the 800x800 simulation as our reference solution and obtain the error by taking the root--mean square difference of the corresponding grid points (i.e. to compare the 200x200 solution to the 800x800 reference solution we extract every fourth point of the reference solution etc.) and normalize relative to the number of grid points. Figure \ref{fig:errorDensity} shows the relative error of density, velocity in x--direction, and pressure as a function of time. 

\begin{figure}
\centering
   \subfloat[error in density, 50 scrt]{\includegraphics[scale=0.3, angle=270]{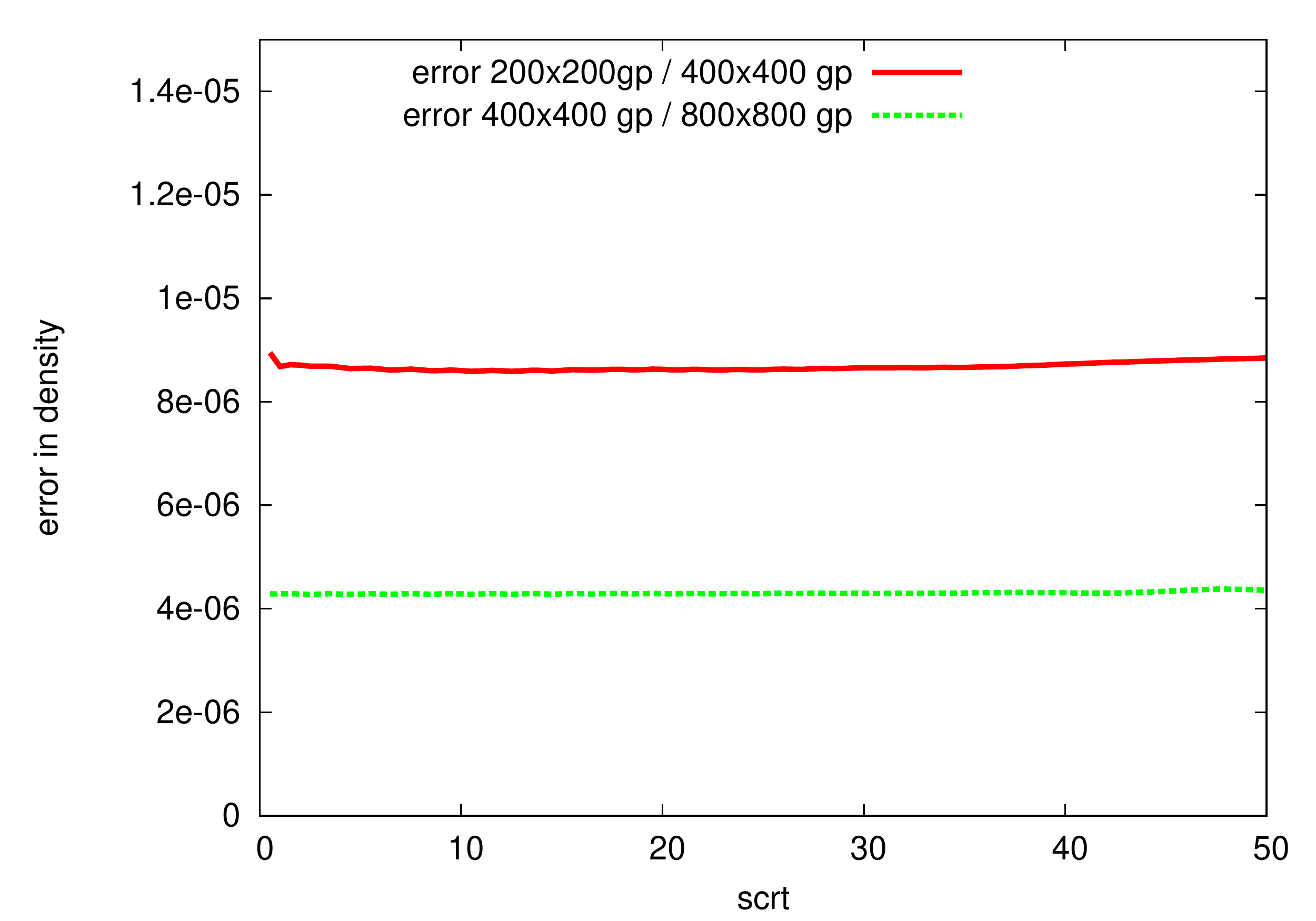}}
   \subfloat[error in density, 200 scrt]{\includegraphics[scale=0.3, angle=270]{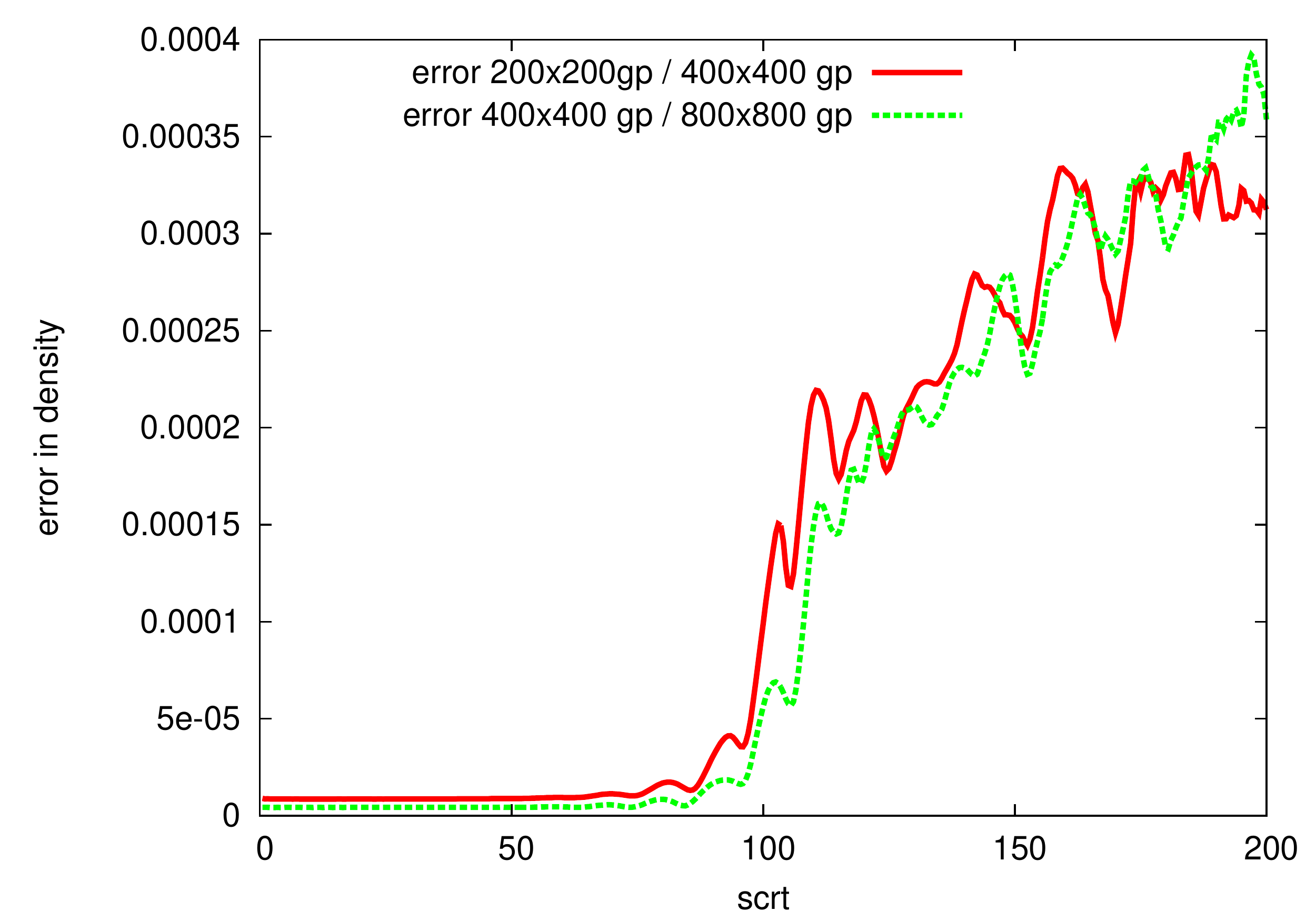}}
   
\centering
   \subfloat[error in x--velocity, 50 scrt]{\includegraphics[scale=0.3, angle=270]{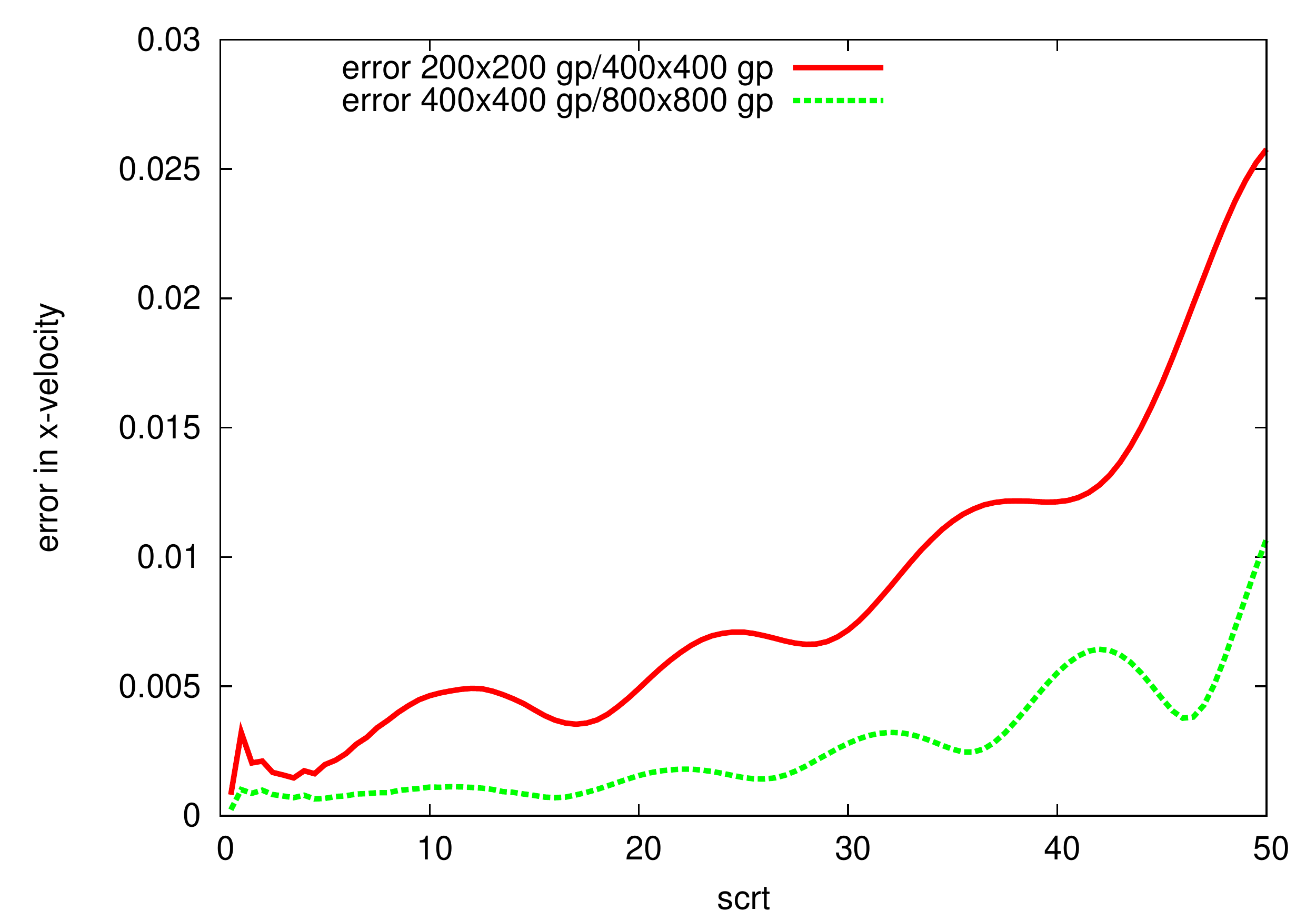}}
   \subfloat[error in x--velocity, 200 scrt]{\includegraphics[scale=0.3, angle=270]{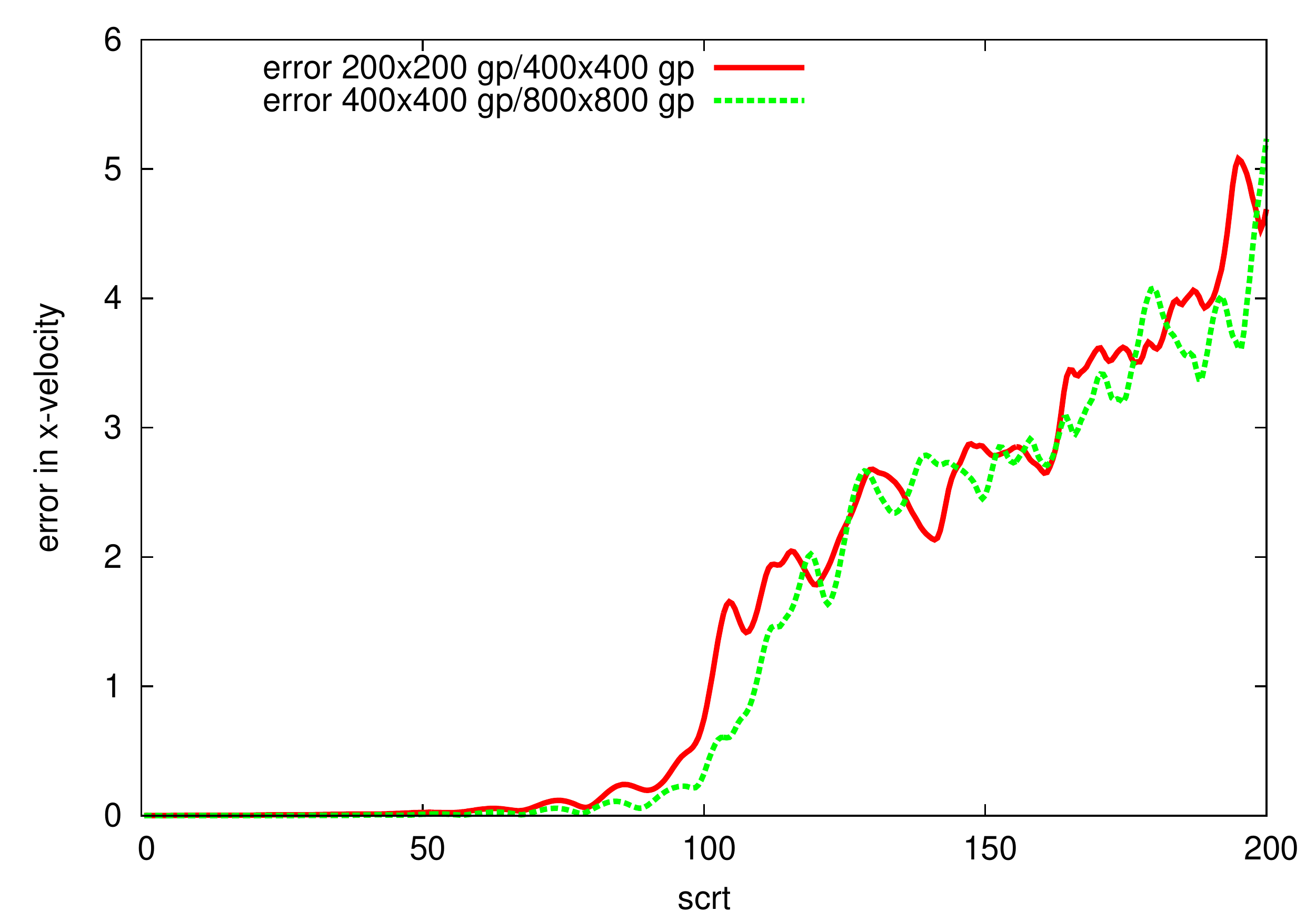}}
   
\centering
   \subfloat[error in pressure, 50 scrt]{\includegraphics[scale=0.3, angle=270]{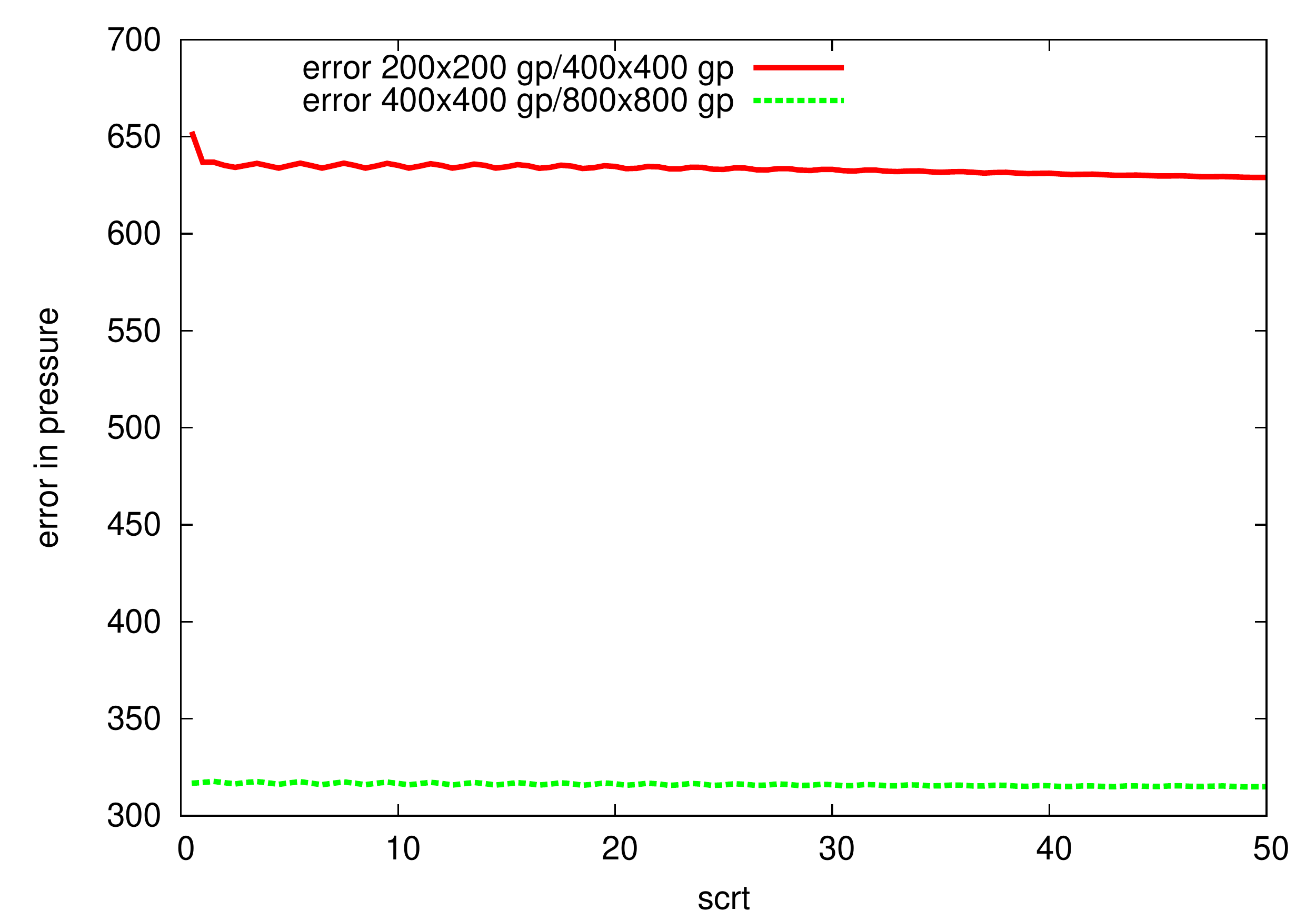}}
   \subfloat[error in pressure, 200 scrt]{\includegraphics[scale=0.3, angle=270]{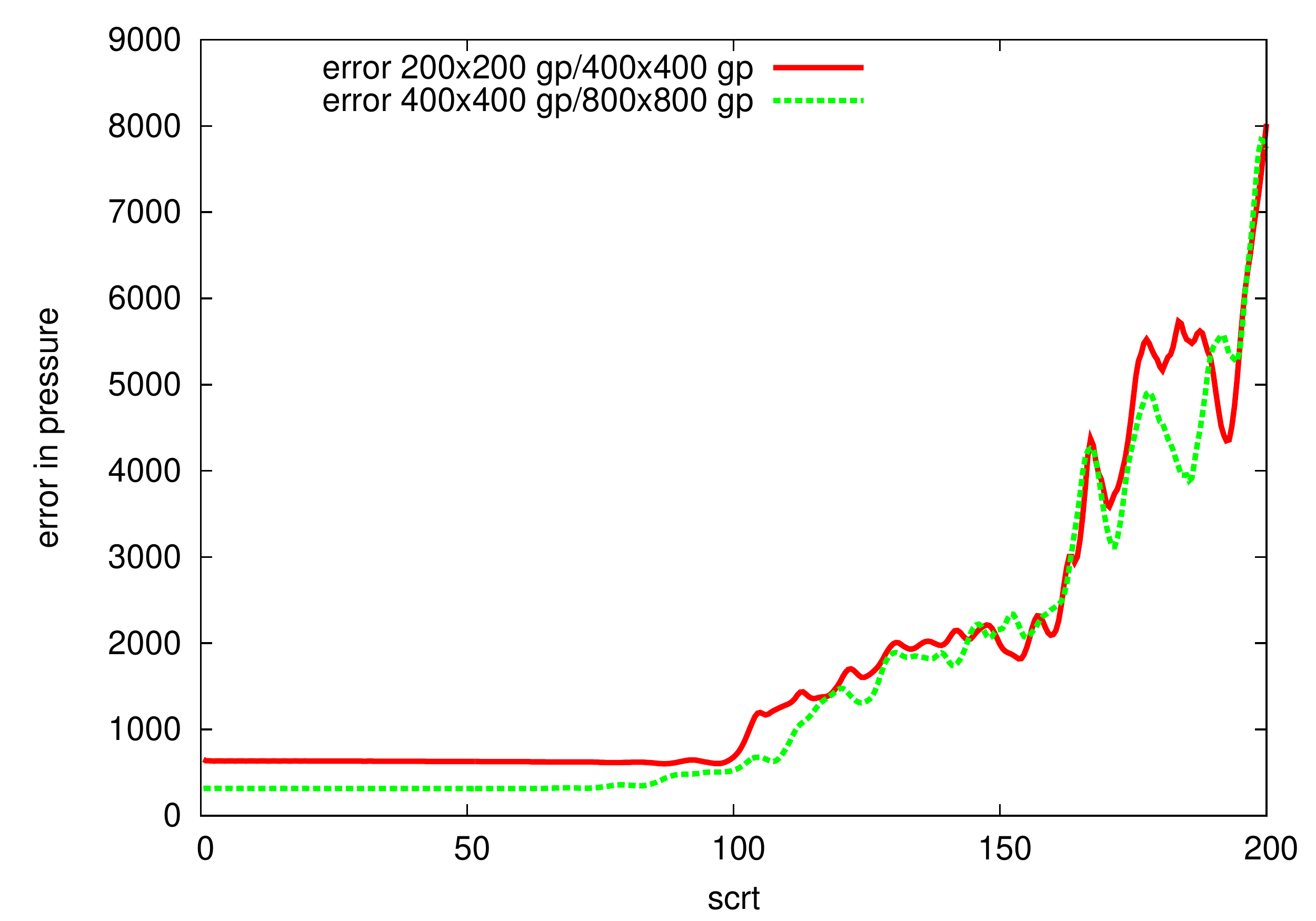}}
   
\caption{Comparison of the error between successively refined grids. }
\label{fig:errorDensity}
\end{figure}

%In the first 50 scrt, where the simulation is dominated by diffusion processes, the relative errors in density and pressure are nearly constant and the error decreases almost quadratically as the resolution increases. 
In the first 50 scrt, where the simulation is dominated by diffusion processes, the relative errors in density and pressure are nearly constant and the error between the simulations with 400$\times$400 and 800$\times$800 grid points is considerably smaller than the error between the simulations with 200$\times$200 and 400$\times$400 grid points. However, a comparison for a given point in time has only limited meaning, since the solution changes its nature as a function of time. Initial vertical oscillations are damped out and the velocity field slowly starts building up. Subsequently, gravity waves form and break. Turbulence sets in and the solution reaches a statistically stationary state. Due to turbulence a direct comparison of the solutions no longer provides meaningful information about the discretization. 

\subsection{Scalability of the fractional step method} \label{sec:scalingtest}
With the advent of high performance supercomputers, efficient parallelization techniques and the scalability of algorithms have become vital for any scientific computation. Although we have already demonstrated the scalability of our solver for the generalized Poisson equation in Section \ref{sec:Poisson}, we would like to demonstrate the scalability of the overall fractional step method, especially, since there are recent publications like \cite{HottaRempel2012} stating that algorithms containing an elliptic equation are limited in parallel efficiency due to the overhead the parallel solution of the elliptic equation incurs. 

To demonstrate that such claims of problematic scalability properties do not hold for the fractional step method, we use the setting of stellar convection with parameters $Pr=0.1$ and $Ra^*=160 000$. In a first test series, we discretize the domain with 1600$\times$1600 grid points and advance the simulation 0.5 soundcrossing times employing a different number of CPUs. Table \ref{kwatraScaling} lists the wallclock times. Indeed, with 64 times the number of cores the code is about 49 times faster which is similar to the scaling reported in Table 2 (for a different computer architecture) for the plain solver for (\ref{eq:GeneralizedPoisson}). Note that the single-core run has not been performed for the entire 0.5 scrt due to the excessively long runtime estimated to be about 7.3 days. However, the overhead of the Schur complement method quantified in Section \ref{sec:Poisson} to be a factor 12 reduces here to a factor of 2.5. This may be due to memory issues -- arrays as large as in the present testcase do not fit easily into the cache memory and a full hydrodynamical simulation requires definitely more memory than the simple solution of the generalized Poisson equation. The emerging latencies due to cache misses considerably slow down the calculation. Clearly, this demonstrates \textit{strong scaling} of the fractional step method over 3 orders of magnitude in number of processor cores. 

The second suite of tests is designed to analyze the behaviour of our solver if more than 1000 cores are used. To reasonably employ each core we use the same setting as in the first test series, but discretize the computational domain with 3200x3200 grid points. The wallclocktimes given in Table \ref{Scaling2} indicate that we have optimal scaling up to 2048 cores. However, employing 4096 cores, there is almost no acceleration compared to the 2048-core run. In order to exclude the solver for the generalized Poisson equation as the cause for this lack of efficiency, we have rerun the same test with the fully explicit solver. The performance of the fully explicit solver is quantified in Table \ref{Scaling3} and again, the employment of 4096 cores does not show a significant advantage over the run with 2048 cores. Therefore, we attribute this lack of acceleration to the specific network structure of the VSC2. Furthermore, with an appropriate parallelization strategy (multigrid solvers for systems of linear equations), corresponding solution techniques are reported to scale up to 290 000 cores \cite{ruede2011}.

Note that due to the exorbitant number of 3200 grid points we use per direction in this test, the dominant timestep restriction stems from diffusive processes. Hence, both solvers use the same time increment and since the solution of the generalized Poisson equation required by the fractional step method does take time, especially, if many grid points are used, the explicit solver is computationally less expensive. We also note here that the number of grid points per domain at 4096 cores is the same as for the test with 1600 grid points per directin on 1024 cores. Thus, the size of the transferred data per domain is the same for both cases, but the total data transfer over the network increases for the larger problem.

\begin{table}[ht!]
 \centering
 \begin{tabular}{|l|l|}
\hline
  \# cores & time in s  \\\hline
  \hline  
  1    & Estimate: 7.3 days \\\hline
  16   &  23:52:50 \\\hline
  64   &  05:53:13 \\\hline
  256  &  01:20:02 \\\hline
  1024 &  00:26:32 \\\hline
 \end{tabular}
\caption{Fractional step method scaling test (1600x1600 grid points), calculated at the Vienna Scientific Cluster 2 (VSC2). }
 \label{kwatraScaling}
\end{table}

\begin{table}[ht!]
 \centering
 \begin{tabular}{|l|l|}
\hline
  \# cores & time in s  \\\hline
  \hline
  256  &  24:15:00 \\\hline
  1024 &  06:39:44 \\\hline
  2048 &  04:44:25 \\\hline
  4096 &  04:38:15 \\
  \hline
 \end{tabular}
\caption{Fractional step method scaling test (3200x3200 grid points), calculated at the Vienna Scientific Cluster 2 (VSC2). }
 \label{Scaling2}
\end{table}

\begin{table}[ht!]
 \centering
 \begin{tabular}{|l|l|}
\hline
  \# cores & time in s  \\\hline
  \hline
  1024 &  02:11:21 \\\hline
  2048 &  01:42:29 \\\hline
  4096 &  01:22:03 \\
  \hline
 \end{tabular}
\caption{Explicit solver scaling test (3200x3200 grid points), calculated at the Vienna Scientific Cluster 2 (VSC2). }
 \label{Scaling3}
\end{table}

\subsection{Validation in the High Mach Number Regime}
%In high Mach number flows, shocks may develop spontaneously and travel through the domain at a certain characteristic speed. <Auswirkungen, Wichtigkeit %von Schocks>. Therefore, the numerical method must be able to capture shocks at the correct speed. 

To validate the shock--capturing capability of the fractional step method, we have rerun the standard Sod shock tube test as outlined in \cite{kwatra}. We use a computational domain of 1 cm and 405 grid points. The initial condition reads
\begin{equation}
(\rho, u, P) =  
\begin{cases}
(1,0,1) & \text{ if $x \leq 0.5 \mathrm{ cm}$} \\
(0.125,0,0.1) & \text{ if $x \geq 0.5 \mathrm{ cm}$}  
\end{cases}
\end{equation} 

\begin{figure}
\centering
   \subfloat[Density]{\includegraphics[scale=0.3, angle=270]{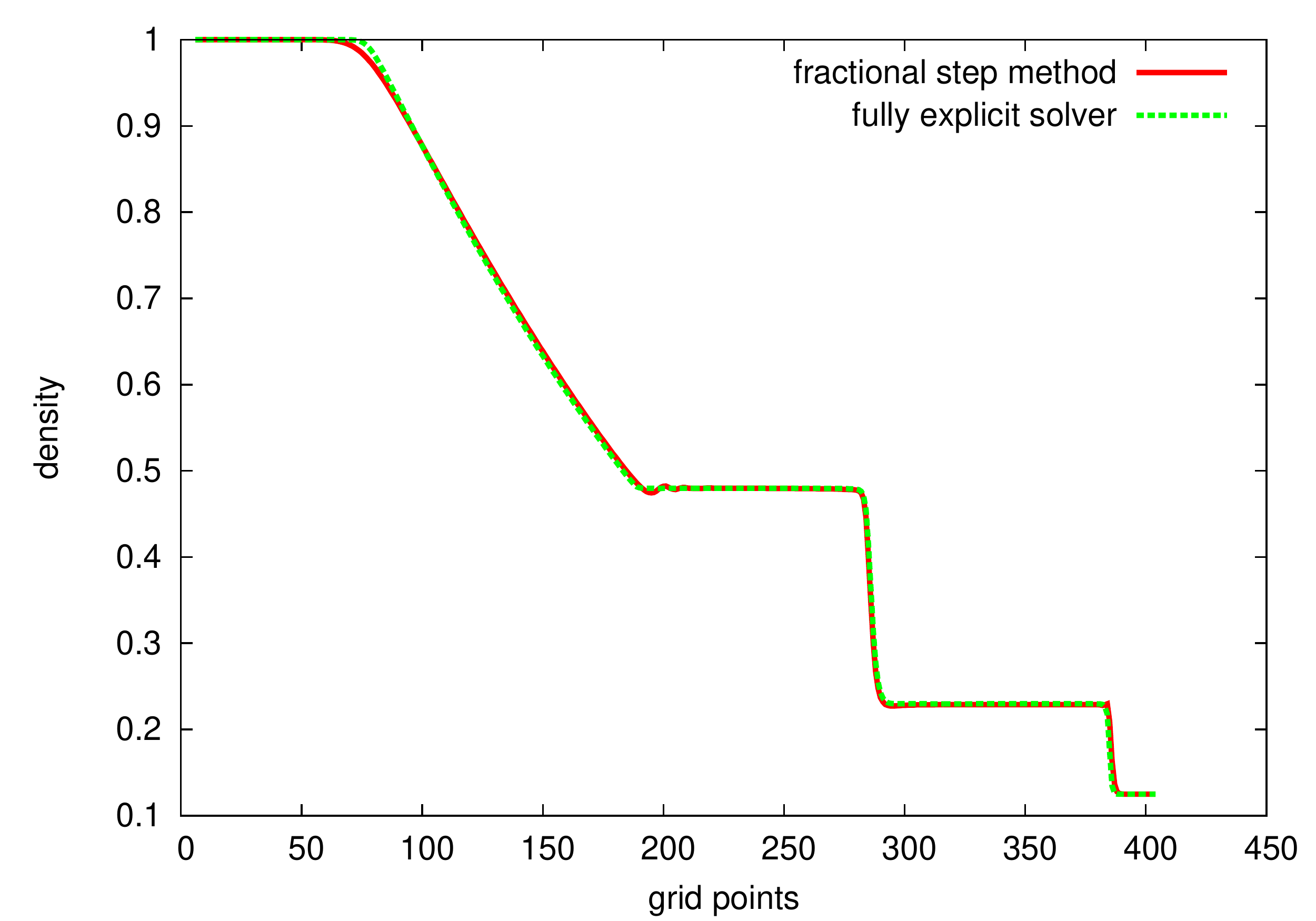}}
   \subfloat[Pressure]{\includegraphics[scale=0.3, angle=270]{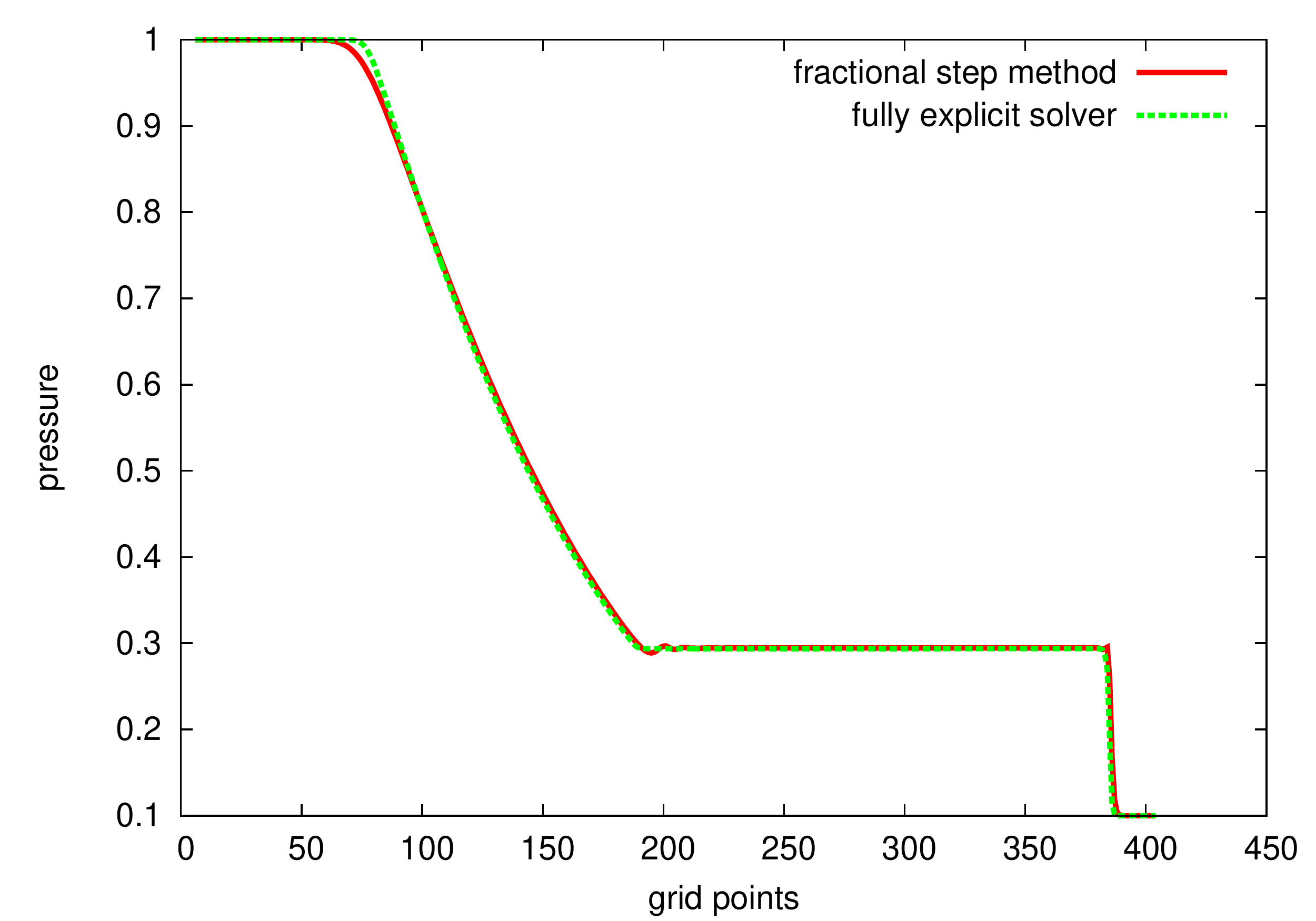}}
\caption{Results of the Sod shocktube test at t=0.25 s. Density and Pressure are plotted against an explicit reference solution. The results indicate well resolved shock waves.}
\label{fig:SodShock1D}
\end{figure}

Figure \ref{fig:SodShock1D} shows a snapshot of the Sod shocktube after $t=0.25 \mathrm{ s}$. The solution of the fractional step method is plotted against an explicit reference solution. The results indicate well resolved shock, rarefaction, and contact solutions travelling at the correct shock speeds. 

We also rerun the two--dimensional circular shock test proposed by \cite{kwatra}.
The initial condition for this test is given by 
\begin{equation}
(\rho, u, v, P) =  
\begin{cases}
(1,0,0,1) & \text{ if $r \leq 40 \mathrm{ cm}$} \\
(0.125,0,0,0.1) & \text{ if $r \geq 40 \mathrm{ cm}$}  
\end{cases}
\end{equation}
 
\begin{figure}[ht!]
  \centering
   \subfloat[Density (semi--implicit)]{\includegraphics[scale=0.18]{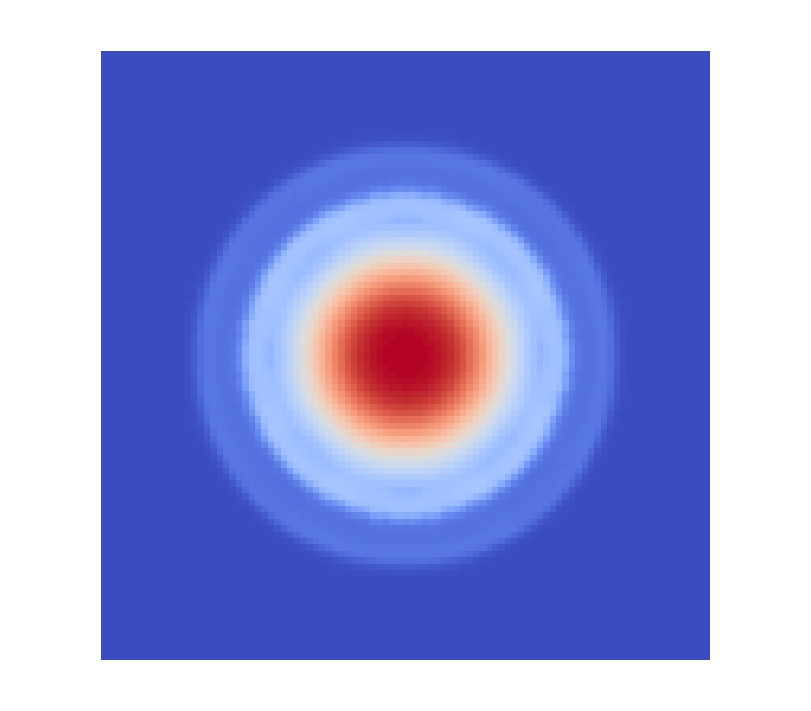}}
   \subfloat[Density (explicit)]{\includegraphics[scale=0.18]{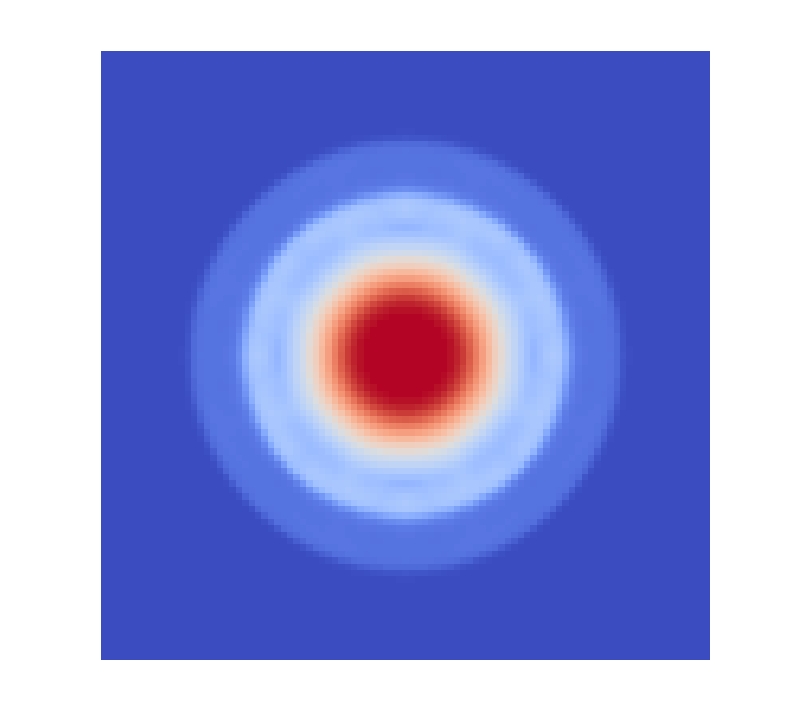}}
   \subfloat[Density]{\includegraphics[scale=0.18]{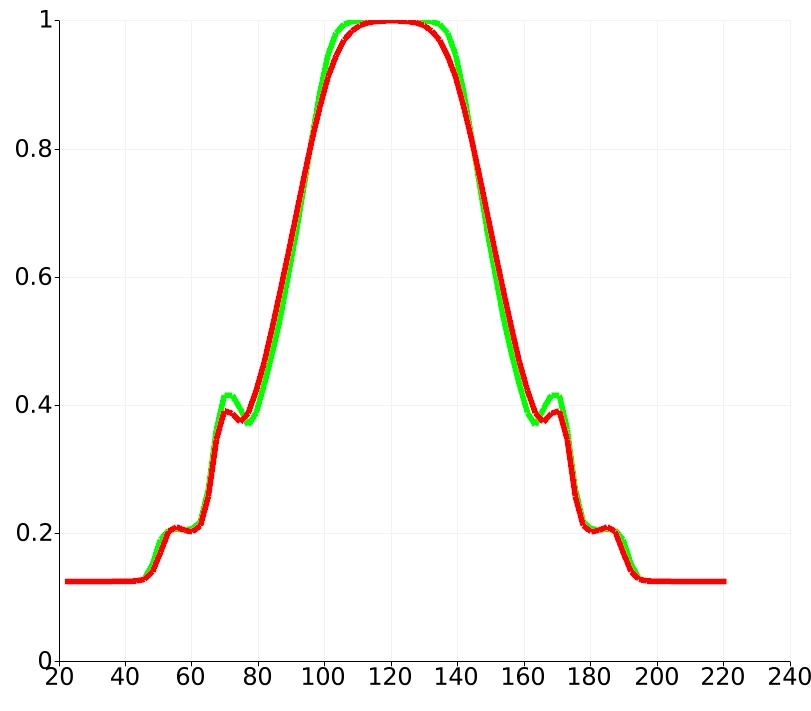}}

\centering
   \subfloat[x--Momentum (semi--implicit)]{\includegraphics[scale=0.18]{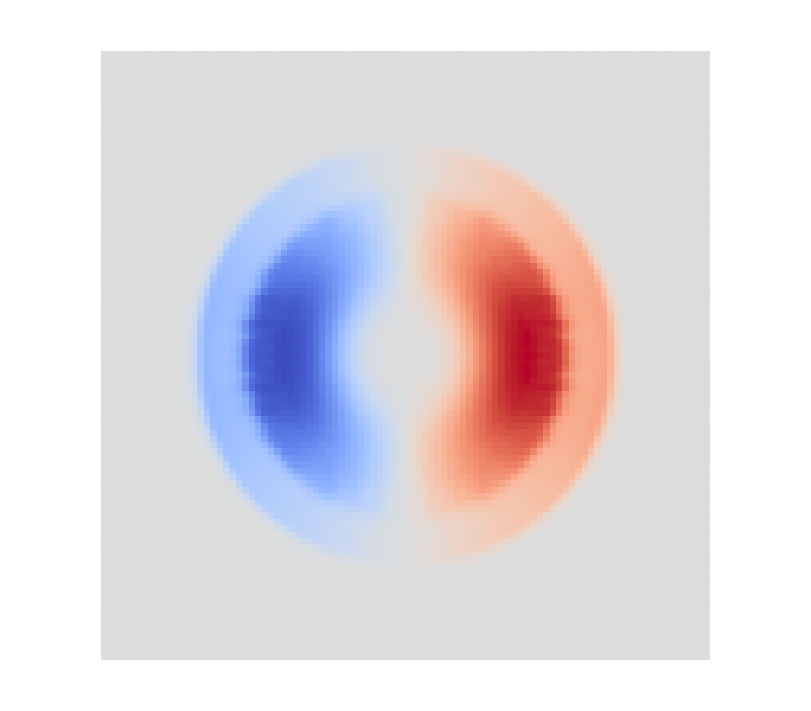}}
   \subfloat[x--Momentum (explicit)]{\includegraphics[scale=0.18]{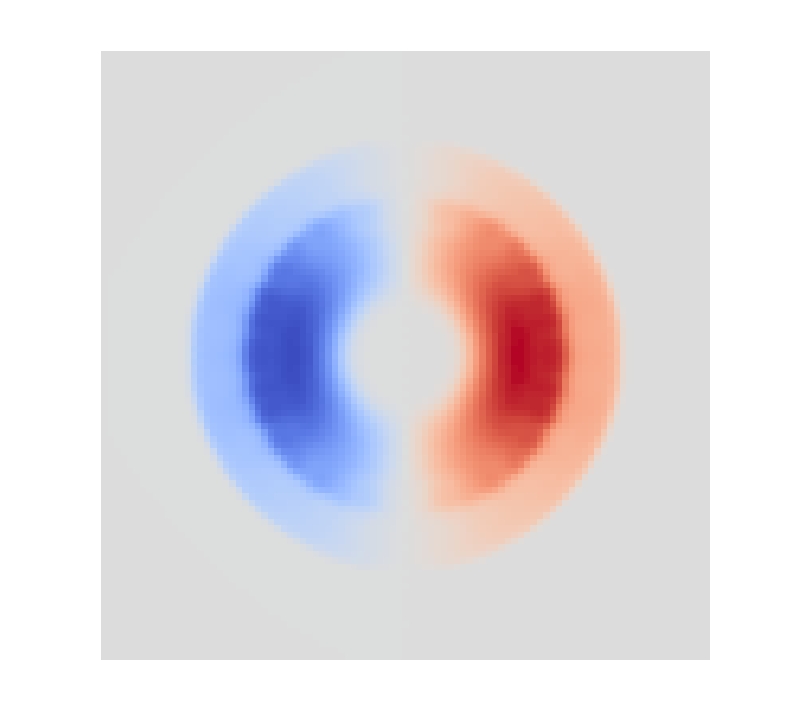}}
   \subfloat[x--Momentum]{\includegraphics[scale=0.18]{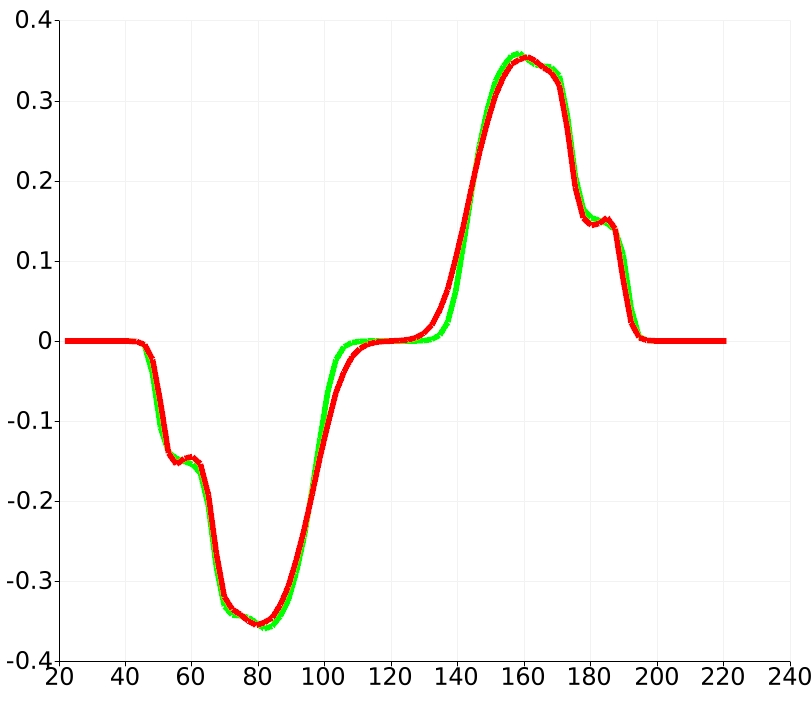}}
\caption{The depicted panels are snapshots of the two--dimensional circular shock test. On the upper left, the density at time t=16.4 s of the simulation performed with the fractional step is shown, next to it the explicitly calculated reference solution. The panel on the right hand side depict a slice of density where the fractional step solution (red) and the fully explicit solution (green) are compared. In the lower part, the same applies to snapshots of the x--momentum.}
\label{fig:CircShock2D}
\end{figure}

We use a domain of 200 cm x 200 cm and discretize it with 95 x 95 grid points. Again, the results depicted in Figure \ref{fig:CircShock2D} indicate well resolved shocks. 

%Note that in this test Mach numbers up to $0.9$ are reached. 

\subsection{Validation in the Low Mach Number Regime}
\subsubsection{Smooth Flow Test}
In \cite{kwatra}, N. Kwatra et al report to have achieved a CFL number of 300 in a smooth flow test with initial conditions

\begin{eqnarray*}
u(x,0) &=& 0 \\
P(x,0) &=& P_0 + \epsilon P_1(x) \\
P_1(x) &=& 6 \cos(2 \pi x) + 10 \sin(4 \pi x) \\
\rho(x,0) &=& \left(\frac{P(x,0)}{\rho_0}\right)^{\frac{1}{\gamma}} \rho_0
\end{eqnarray*}

$\gamma$ is the adiabatic index of the gas and $\rho_0 = 1$, $P_0 = 10^{3}$, $\epsilon = 1$.

Contrary to \cite{kwatra}, we use a fixed grid resolution of 800 grid points and increase the Courant number. The timing results in Table \ref{SFT-times} show that the semi--implicit method is far more efficient than the explicit method even when the same time increment is used. This was already remarked by \cite{kwatra} and predominantly attributed to the avoidance of the characteristic decomposition in the ENO scheme. However, in the scaling test in Section \ref{sec:scalingtest} quite the opposite was true, namely, the explicit solver was considerably faster than the semi-implicit method when the same timestep was employed. We attribute this to the number of grid points - in the smooth flow test, we discretize the domain with 100 grid points whereas in the scaling test we employed 3200x3200 grid points. Apparently, as the number of grid points increases, the gain in efficiency due to the avoidance of the transformation onto the characteristic space is cancelled by the time spent in the solver for the generalized Poisson equation. 
Note that in the Smooth Flow test, the maximum time increment is 600 times that one of the explicit method, its wallclocktime 234 times smaller than the explicit method and 105 times smaller than the semi-implicit method at the same small time-increment. The loss of a factor of about 6 in efficiency is due to the larger number of iterations during the solution of (\ref{eq:GeneralizedPoisson}) needed for the  much larger time-steps. We also remark that although it is possible to employ a CFL-number of 300 without running into numerical difficulties it is not advisable to do so for long integration times since the dissipation added by the WENO-scheme damps the original flow pattern considerably. 

\begin{figure}[ht!]
  \centering
   \includegraphics[scale=0.35, angle=270]{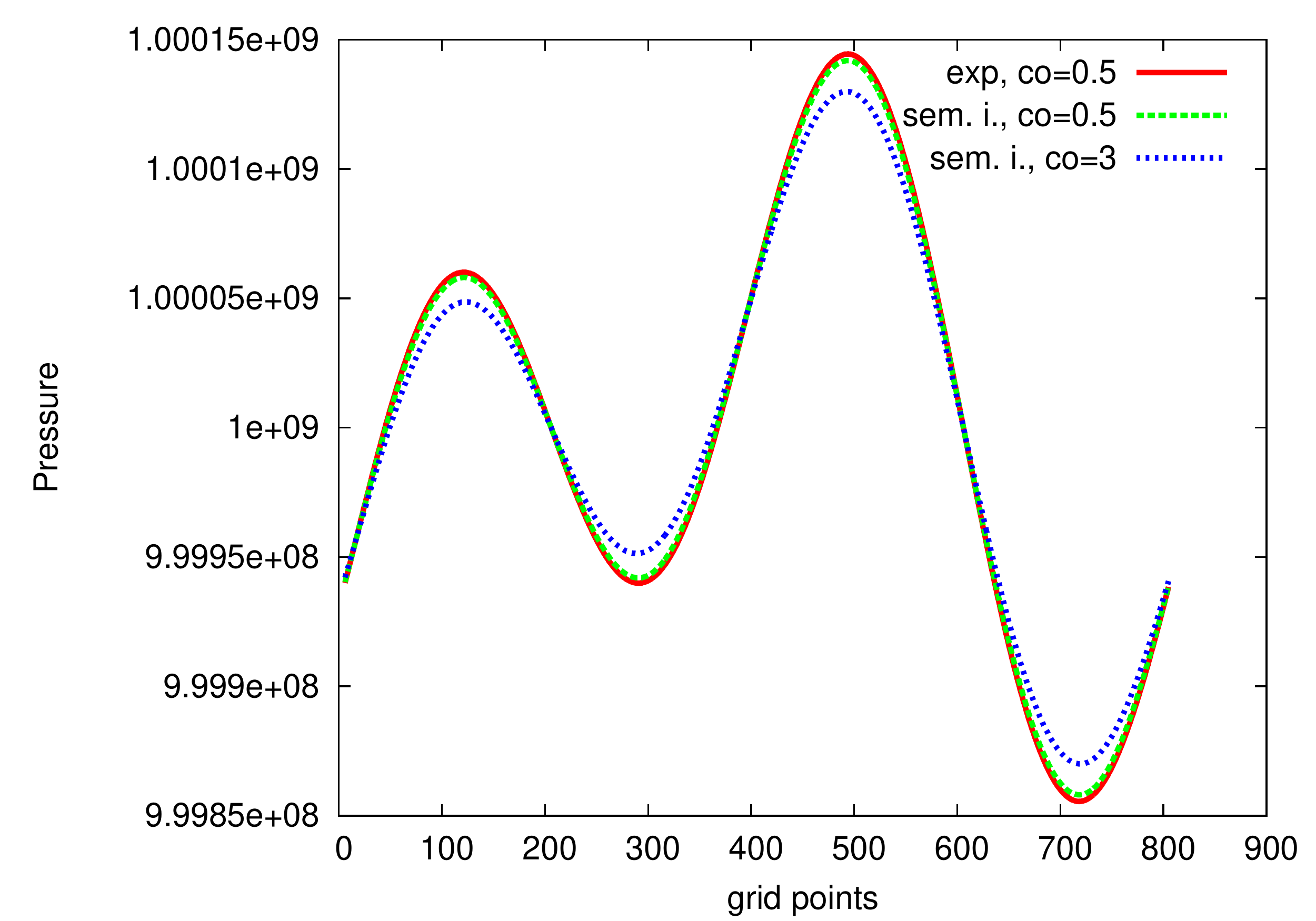}
\caption{Numerical results comparing the pressure at t=$1.25\times 10^{-5}$ s in the smooth flow test run with different CFL numbers.}
\label{fig:SFT}
\end{figure}

Figure \ref{fig:SFT} depicts snapshots of pressure in the smooth flow test run at different Courant numbers. 

\begin{table}[ht!]
 \centering
 \begin{tabular}{|l|l|l|l|}
\hline
  method & CFL-number & timestep $\Delta t$ & Wallclocktime  \\\hline
  \hline
  explicit & 0.5 & $1.53\times10^{-8}$ & 01:18:04 \\ \hline
  semi-implicit & 0.5 &$1.53 \times 10^{-8}$ & 00:35:04 \\ \hline
  semi-implicit & 3 &$9.18\times10^{-8}$ & 00:06:33 \\ \hline
  semi-implicit & 30 &$9.18\times10^{-7}$ & 00:01:59 \\ \hline
  semi-implicit & 300 &$9.18\times10^{-6}$ & 00:00:20 \\ \hline
  \hline
 \end{tabular}
\caption{Timing results from the smooth flow test for different Courant numbers. Simulation time is t=$2.5\times 10^{-5}$ s.}
 \label{SFT-times}
\end{table}

\subsubsection{The Gresho Vortex test}

The \textit{Gresho vortex} is a time-independent rotation pattern. Angular velocity depends only on the radius and centrifugal force is balanced by the 
pressure gradient. The original setup is found in \cite{liskawendroff2004}. We use the slightly modified initial condition of \cite{FMiczek2011}, which permits the variation of the Mach number. 

We use a Cartesian domain [0,1] $\times$ [0,1] and employ periodic boundary conditions. The initial condition is given dependent on the radius 
\sloppy $ r = \sqrt{(x-0.5)^2 + (y-0.5)^2}$ as

\begin{eqnarray}
\rho &=& 1.0 \\
P_0 &=& \frac{\rho}{\gamma Ma^2} \\
u_{\phi} &=& 
\begin{cases}
5r & \text{ if $0 \leq r \leq 0.2$} \\
2-5r & \text{ if $0.2 \leq r \leq 0.4$} \\
0 & \text{ if $0.4 \leq r$}  
\end{cases} \\
P &=& 
\begin{cases}
P_0 + \frac{25}{2}r^2 & \text{ if $0 \leq r \leq 0.2$} \\
P_0 +  \frac{25}{2}r^2 + 4\; (1-5r-\mathrm{ln}(0.2) + \mathrm{ln}(r)) & \text{ if $0.2 \leq r \leq 0.4$} \\
P_0 -2+4\;\mathrm{ln}(2) & \text{ if $0.4 \leq r$}  
\end{cases}
\end{eqnarray}
 
$Ma$ denotes the Mach number and $u_{\phi}$ denotes the angular velocity. The cartesian velocity components are obtained via 

\begin{eqnarray}
u_x &=& \sin(\theta) \; u_{\phi} \\
u_y &=& \cos(\theta)\; u_{\phi}
\end{eqnarray}

where $\theta = \mathrm{atan2}(y-0.5, x-0.5)$. 

\vspace{1em}

\begin{figure}[ht!]
  \centering
   \subfloat[Ma=0.1, t=0 s]{\includegraphics[width=5cm, height=4cm]{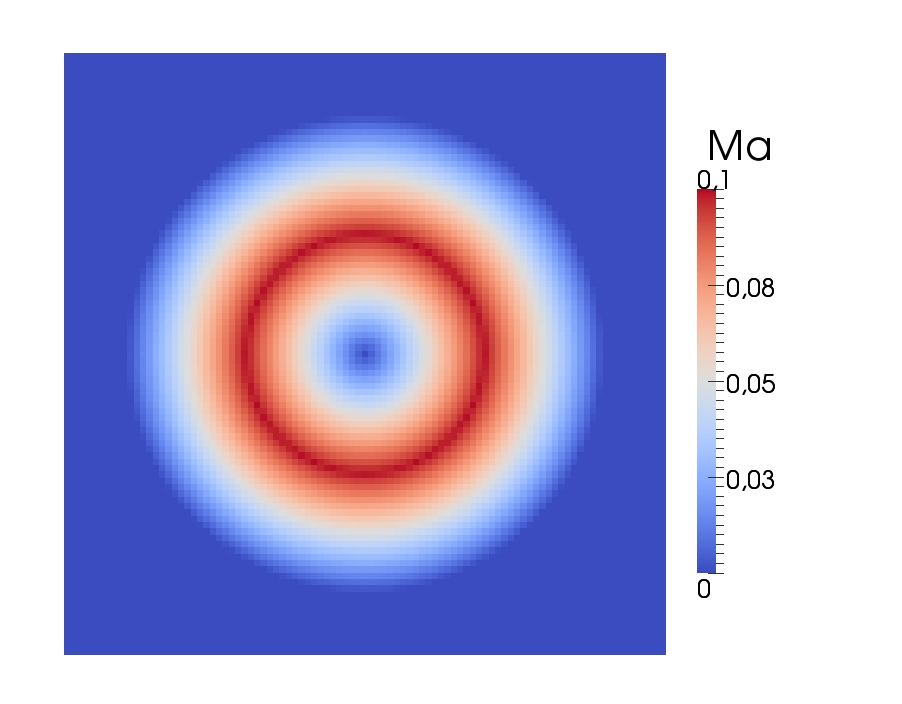}}
   \subfloat[Ma=0.1, t=2 s, exp.]{\includegraphics[width=5cm, height=4cm]{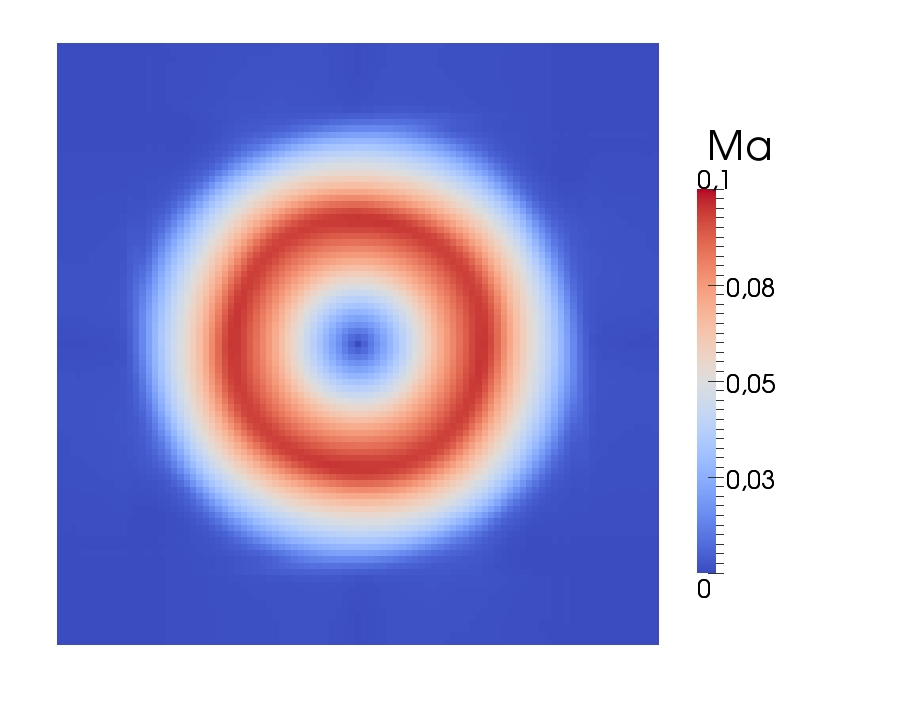}}
   \subfloat[Ma=0.1, t = 2 s, sem.i.]{\includegraphics[width=5cm, height=4cm]{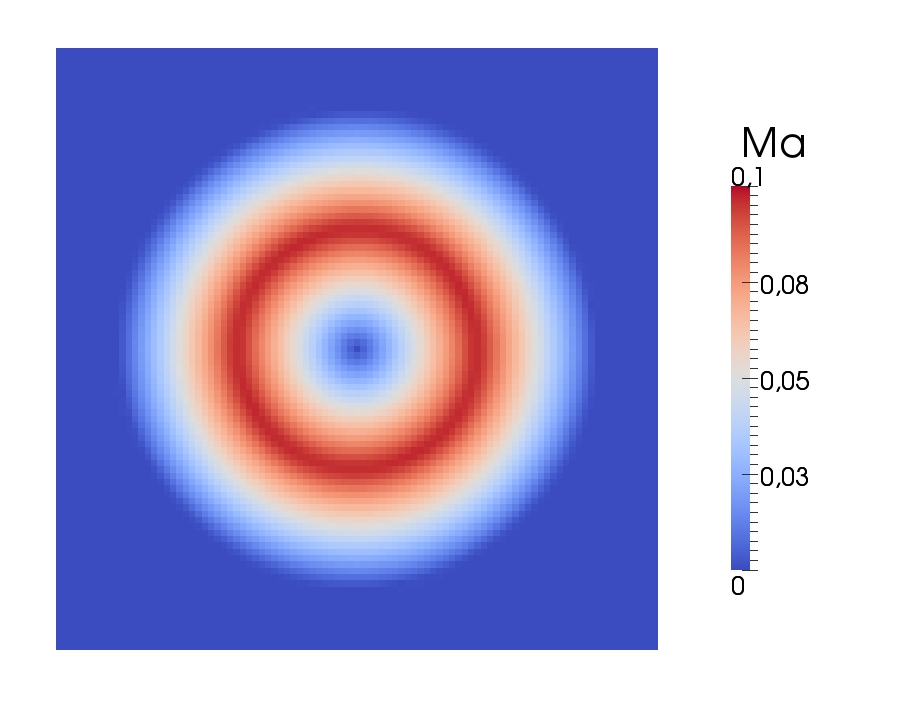}}

\centering
\subfloat[Ma=0.01, t=0 s]{\includegraphics[width=5cm, height=4cm]{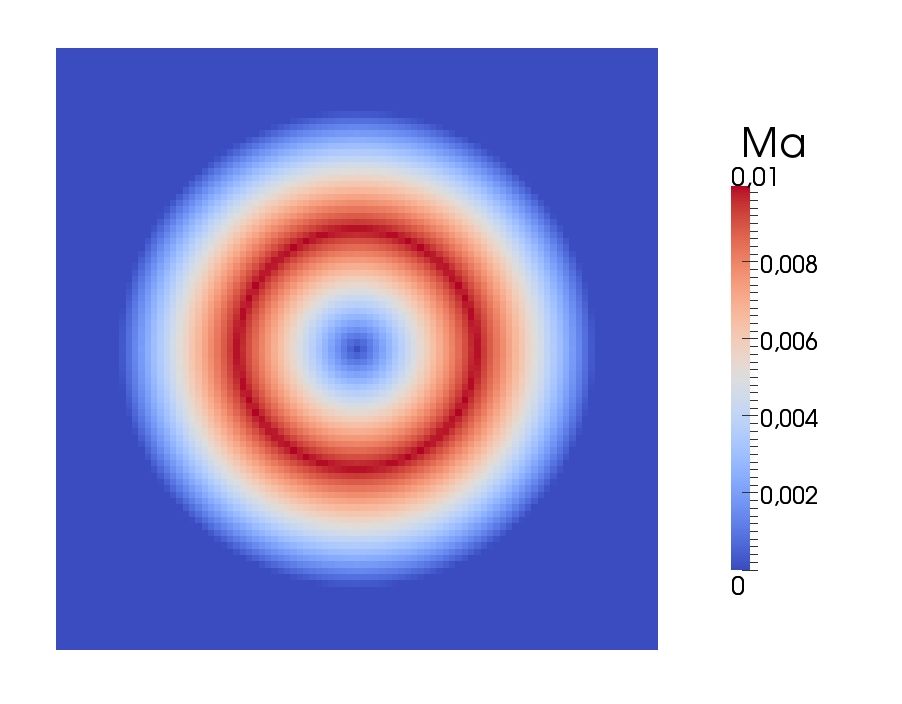}}
   \subfloat[Ma=0.01, t=2 s, exp.]{\includegraphics[width=5cm, height=4cm]{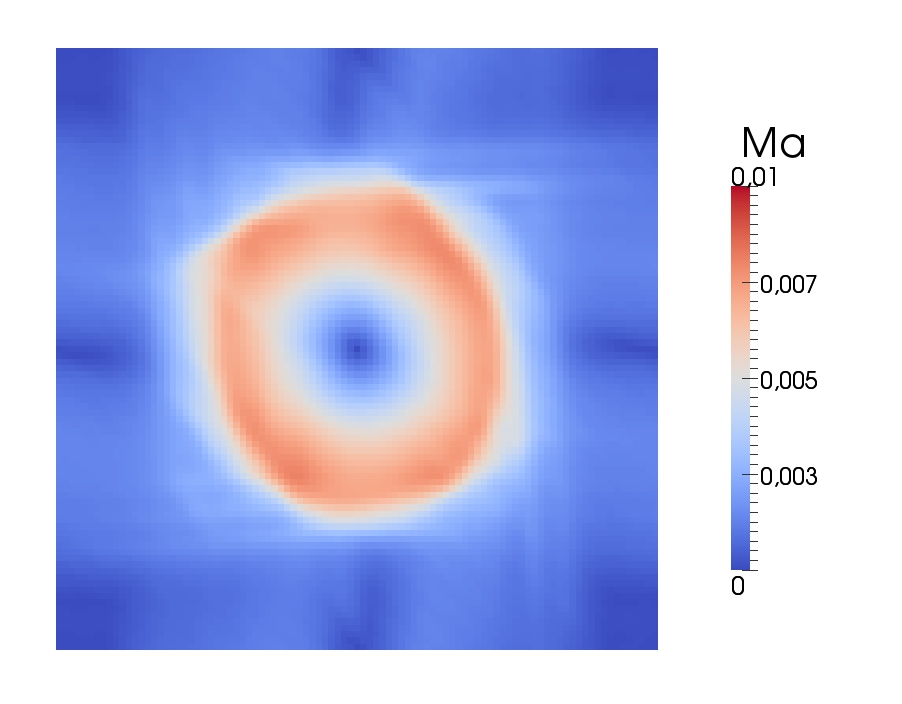}}
   \subfloat[Ma=0.01, t=2 s, sem.i.]{\includegraphics[width=5cm, height=4cm]{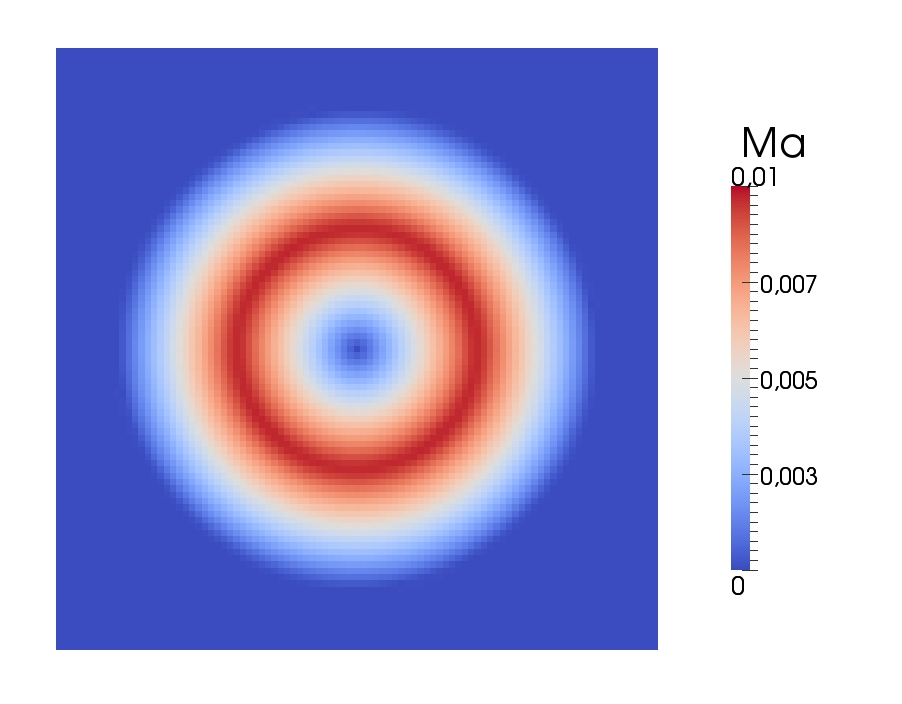}}

\centering
   \subfloat[Ma=0.001, t=0 s]{\includegraphics[width=5cm, height=4cm]{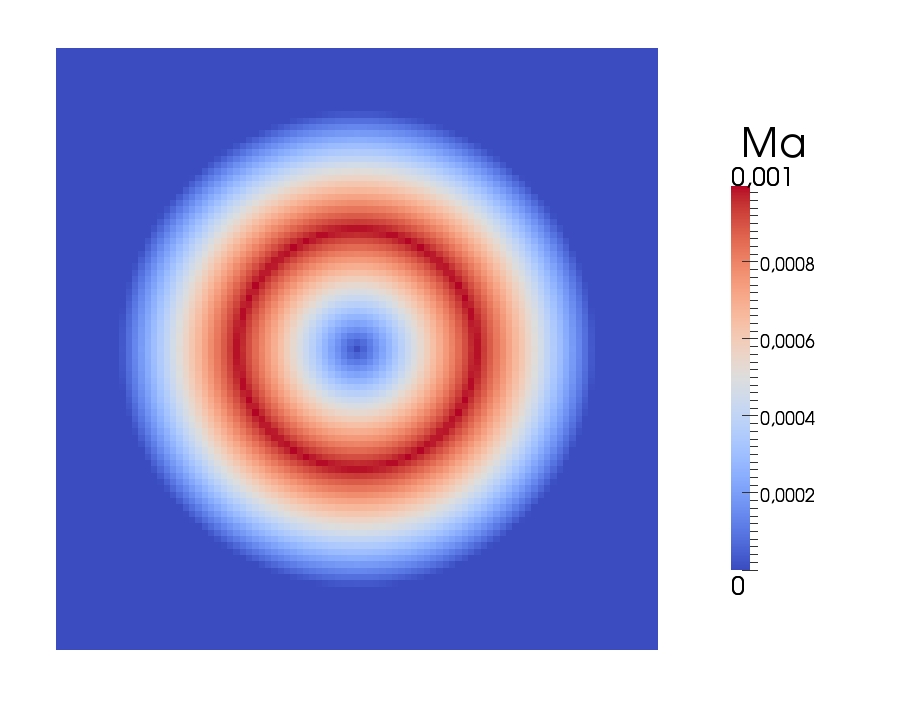}}
   \subfloat[Ma=0.001, t=2 s, exp.]{\includegraphics[width=5cm, height=4cm]{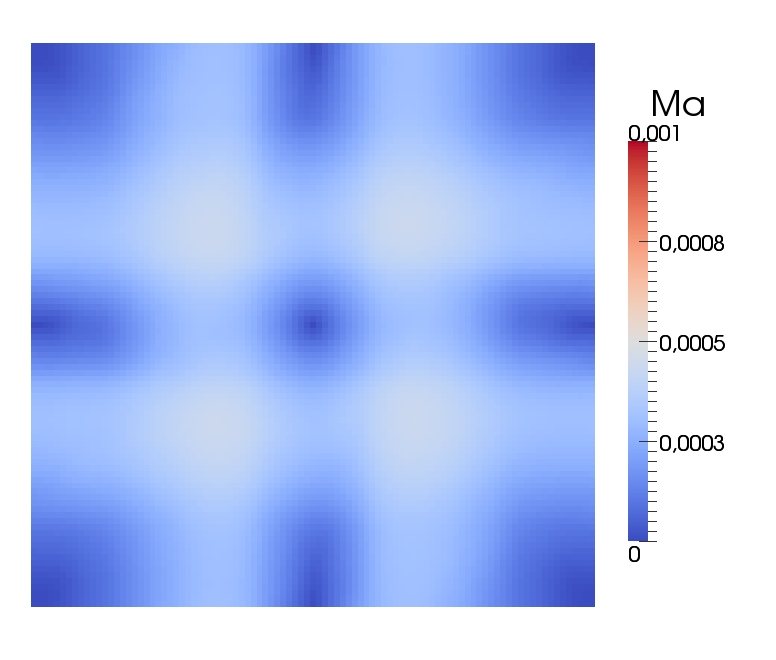}}
   \subfloat[Ma=0.001, t=2 s, sem.i.]{\includegraphics[width=5cm, height=4cm]{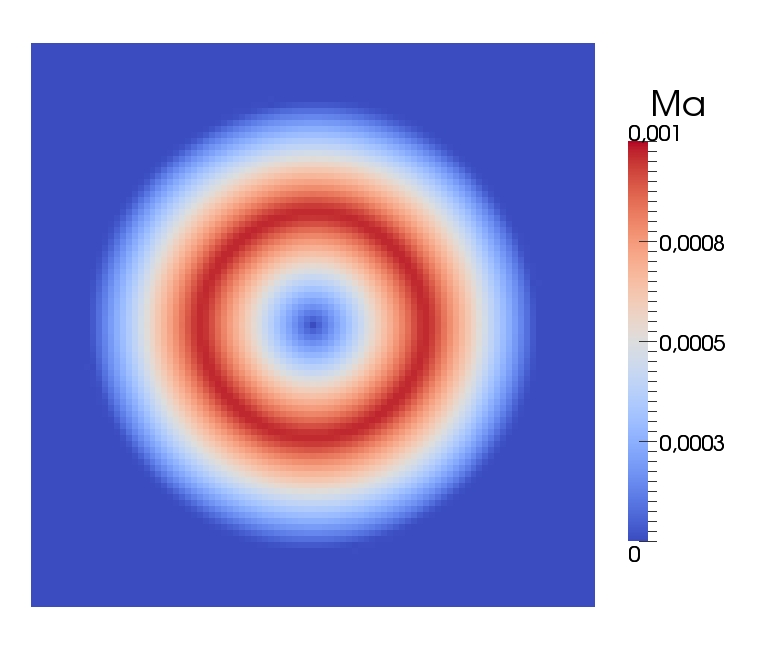}}
   
   \caption{Comparison of the Gresho vortex test performed with the explicit and the semi--implicit solver with different Mach numbers. On 
the left, the initial configuration is depicted. In the center and on the right, the simulation is advanced 2 sec with the explicit solver and the semi--implicit solver, respectively.}
   \label{fig:CompGresho}        
\end{figure}

We run the Gresho vortex test with $Ma=0.1$, $Ma=0.01$ and $Ma=0.001$. Figure \ref{fig:CompGresho} compares the results obtained in this setting by the 
semi--implicit method and the fully explicit solver. In this test, both solvers use, for proper comparison, the same sound--speed induced timestep and a Courant 
number of 0.5. The simulation time is 2 sec. 

\begin{figure}[ht!]
\centering
  	\includegraphics[scale=0.3, angle=270]{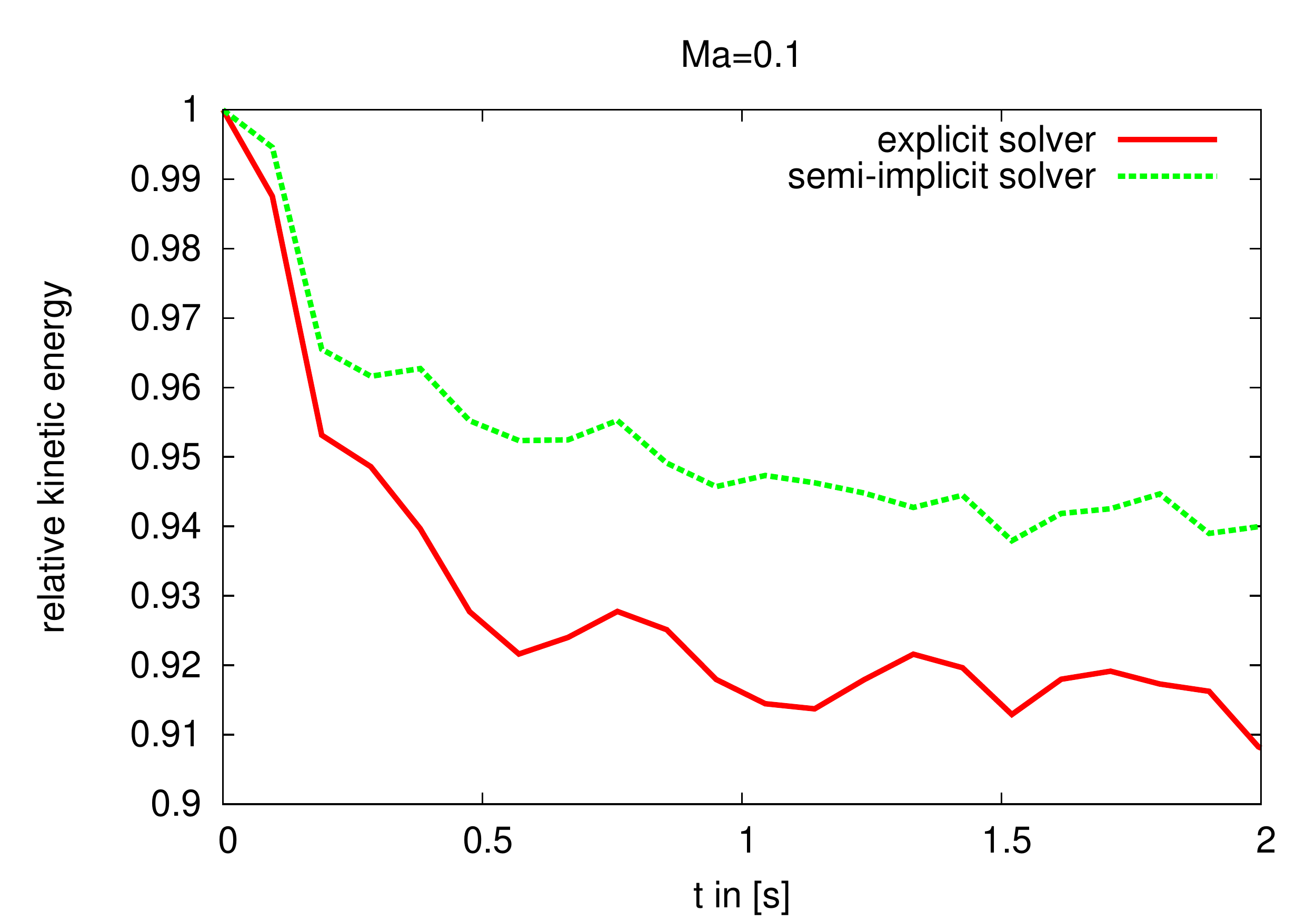}
  	
\centering
  	\includegraphics[scale=0.3, angle=270]{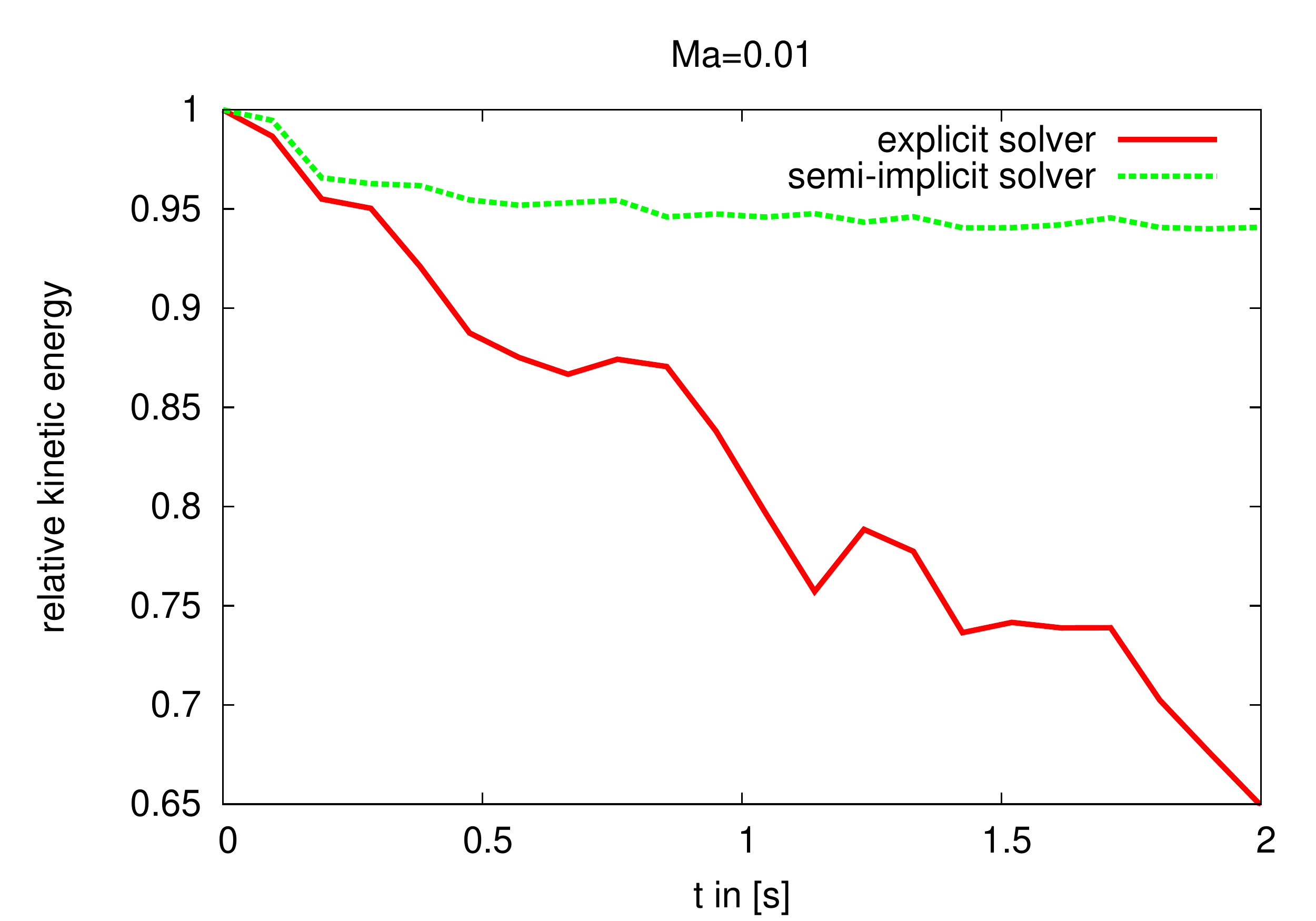}

\centering
  	\includegraphics[scale=0.3, angle=270]{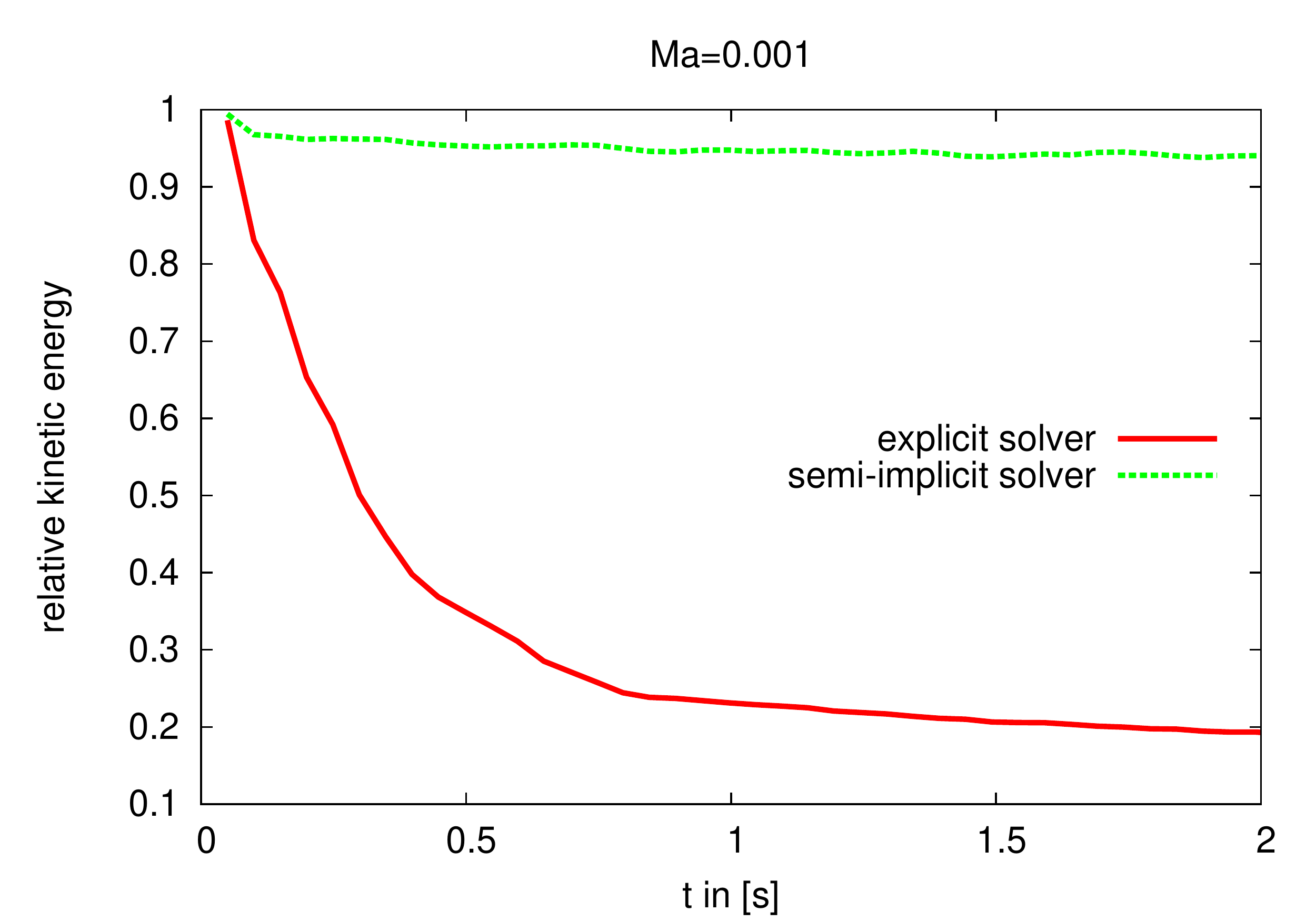}
  	
   \caption{We compare the relative kinetic energy predicted by the simulations over time. Whereas the dissipation of the explicit solver increases 
dramatically as the Mach number drops, the dissipation of the semi--implicit method is independent of the Mach number for exactly the same number of timesteps. } 
        \label{fig:dissipation1}  
\end{figure}

The results show that for $Ma=0.1$ both solvers yield similar results, although the dissipation of kinetic energy is higher in the explicit solver 
(see also Figure \ref{fig:dissipation1}). However, as the Mach number drops, the solution of the explicit solver becomes highly inaccurate. This breakdown is well known in the literature and attributed to  the fact that the pressure term is of order $1/Ma^2$, which introduces considerable inaccuracy as the Mach number approaches 0 (see \cite{GuillardMurrone2003,LowMachReview,zaussinger}). However, in \cite{KarkiPatankar1989} it is shown that  within pressure based methods such as the fractional step method of Kwatra et al. \cite{kwatra} the pressure variation remains finite, irrespectively of the Mach number. This is vividly demonstrated by the highly accurate results of the semi--implicit method even at Mach numbers as low as $10^{-3}$.  
 
\begin{figure}[t]
\centering
\includegraphics[scale=0.3, angle=270]{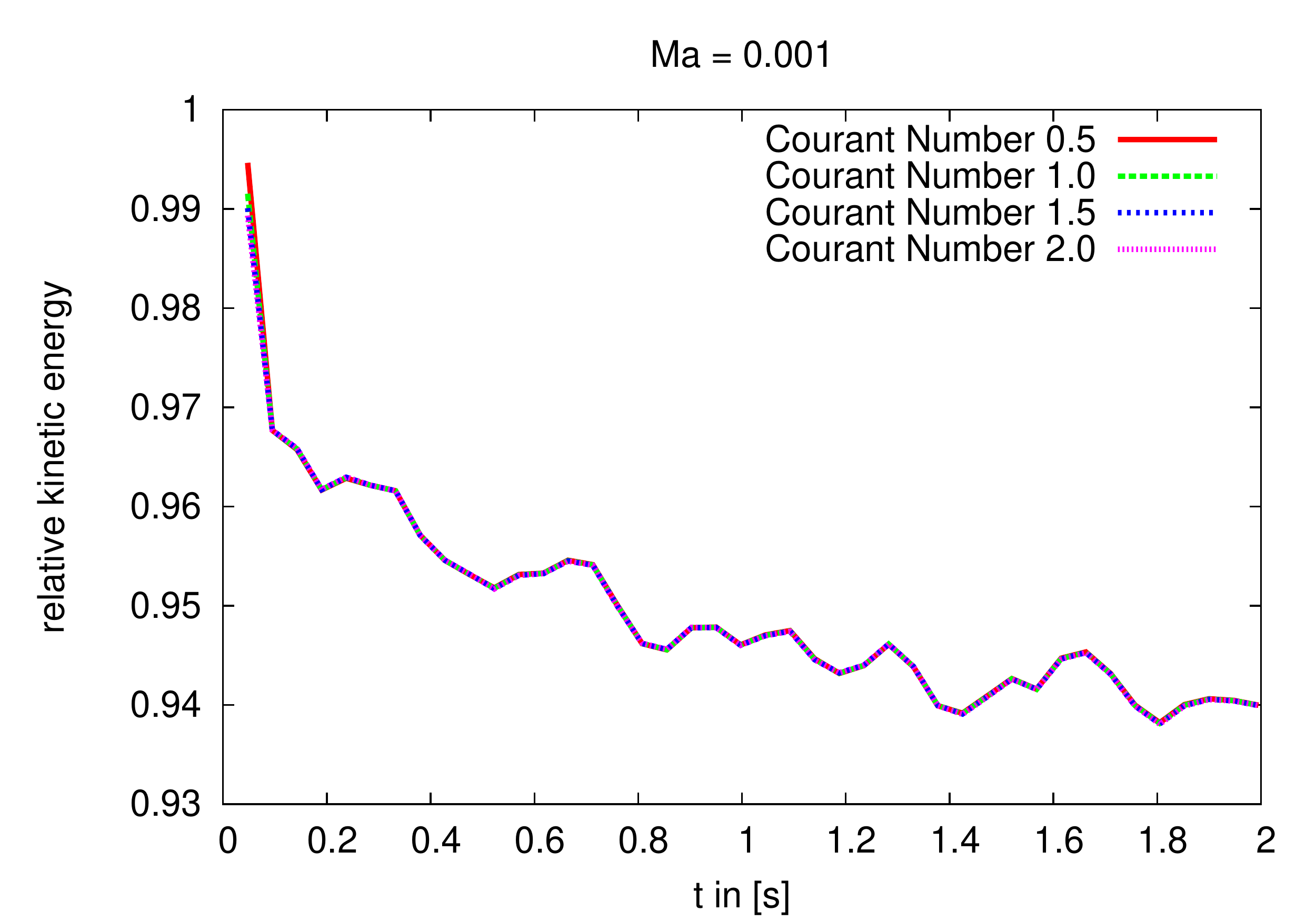}
   \caption{Courant number tests with the semi-implicit method. Advancing the Gresho vortex with $Ma=0.001$ with a greater timestep does not affect the dissipation of the scheme.}
   \label{fig:dissipation2}
\end{figure}

Since the semi-implicit method theoretically allows the simulation to be advanced with a timestep higher than the one induced by the sound speed, we have rerun the Gresho vortex test with $Ma=0.001$ and the semi-implicit solver, advancing it by multiples of the sound-speed induced timestep. We find that the simulation remains stable for up to four times the explicit timestep. Figure \ref{fig:dissipation2} shows no adverse effects on the dissipation from choosing a larger timestep.

\section{Conclusions and Outlook}
We have extended the numerical method of \cite{kwatra} to viscous two-component flows and improved the stability of the overall scheme by employing a strictly dissipative finite difference scheme for the parabolic part. We have applied it to simulations of stellar convection and stellar semiconvection, where especially the latter is situated in the low Mach number regime. Apart from rendering highly accurate results, the implicit treatment of sound waves allows the employment of a much larger timestep to advance the simulations in time. Contrary to most solvers designed for low Mach number regimes, this method is capable of passing smoothly to higher Mach numbers. This makes it especially suited to problems, where a wide range of parameter sets is covered. A prominent example of such a setting is the solar convection zone, where at the bottom low Mach numbers of order $O(10^{-4})$ are encountered, whereas at the top of the same convection zone Mach numbers of order $O(1)$ are reached at least locally inside fast downdrafts, (see \cite{cobold2011,kupkaLecPhys}).  

We have also validated the shock capturing capability of the fractional step method as well as its performance in very low Mach number regimes. We find that the fractional step method renders stable and accurate results even in regimes with $Ma=0.001$. 

By employing a highly parallelized solver for the generalized Poisson equation, we show that the method is suitable for modern high performance computing. We demonstrate the strong scaling of the method over three orders of magnitude in number of processor cores in a realistic astrophysical application. 

Given the beneficial properties of this low Mach number solver outlined in this paper, we conclude that it poses a real alternative to traditional low Mach number schemes such as the Boussinesq or the anelastic approximation which are unreliable in the transition region of moderate Mach numbers (Ma $\approx$ 0.1) and violate various conservation properties (mass in the case of the Boussinesq approximation and energy in case of the anelastic approximation of \cite{randall2010}) 

%We have tested the numerical method of \cite{kwatra} in the low Mach number limit as well as its performance in high Mach number flows. Apart from rendering highly accurate results in the very low Mach number regime, it allows simulations to be advanced with a timestep significantly higher than a fully explicit solver and contrarily to most solver designed for simulations of that kind, it  is capable of passing 
%smoothly to regimes with higher fluid velocities and even handles shock waves correctly. Thus it is applicable to a wide range of problems and especially suited to problems where a wide range of parameter sets is covered. 

%Furthermore, we have generalized this approach to the more complex Navier-Stokes equations and successfully enhanced the stability of the overall scheme, 
%making it thus suitable to more realistic long-term simulations of astrophysical processes. 

\vspace{1em}

\begin{figure}
\centering
\includegraphics[width=5cm, height=8cm, angle=270]{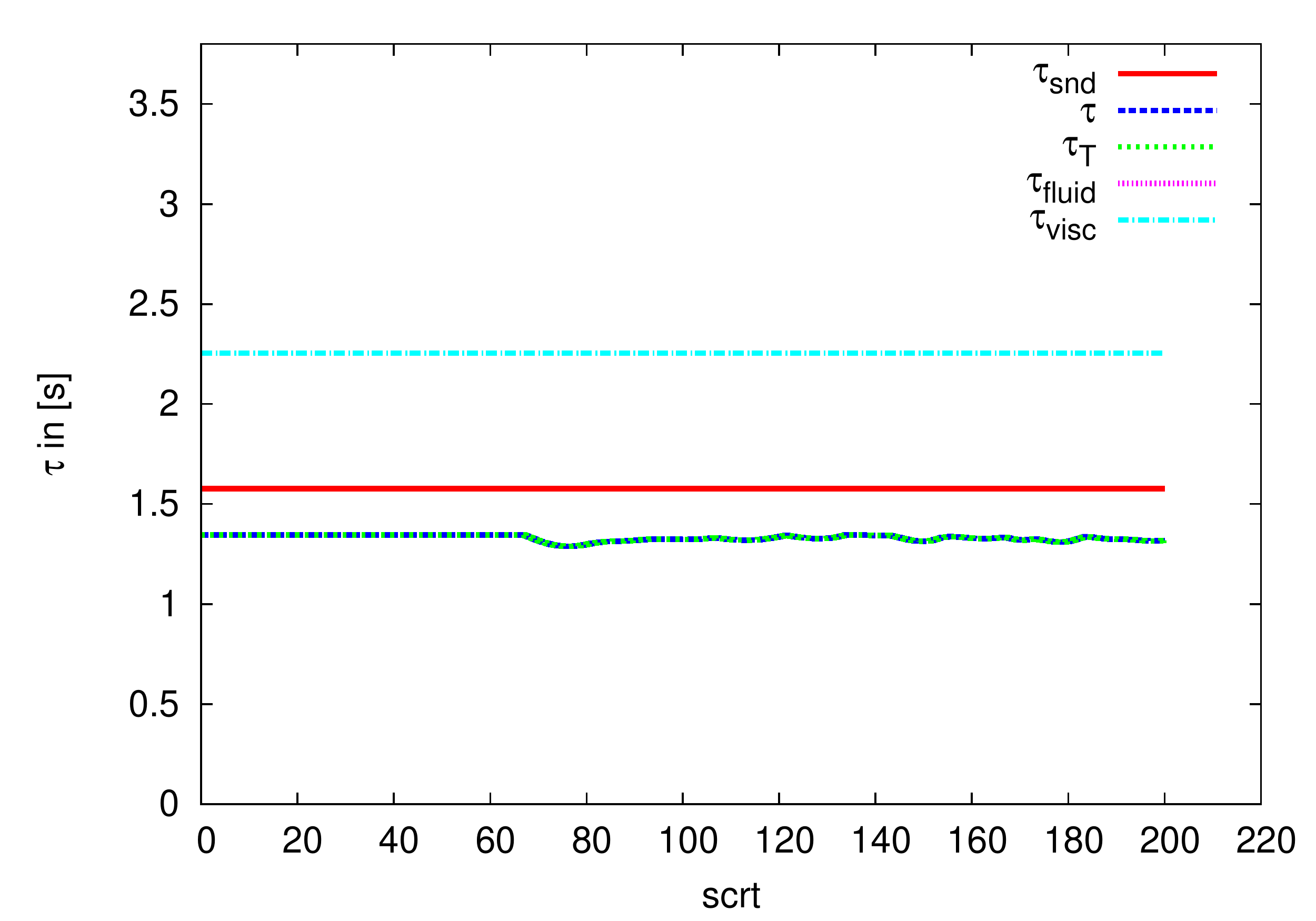}
\includegraphics[width=5cm, height=8cm, angle=270]{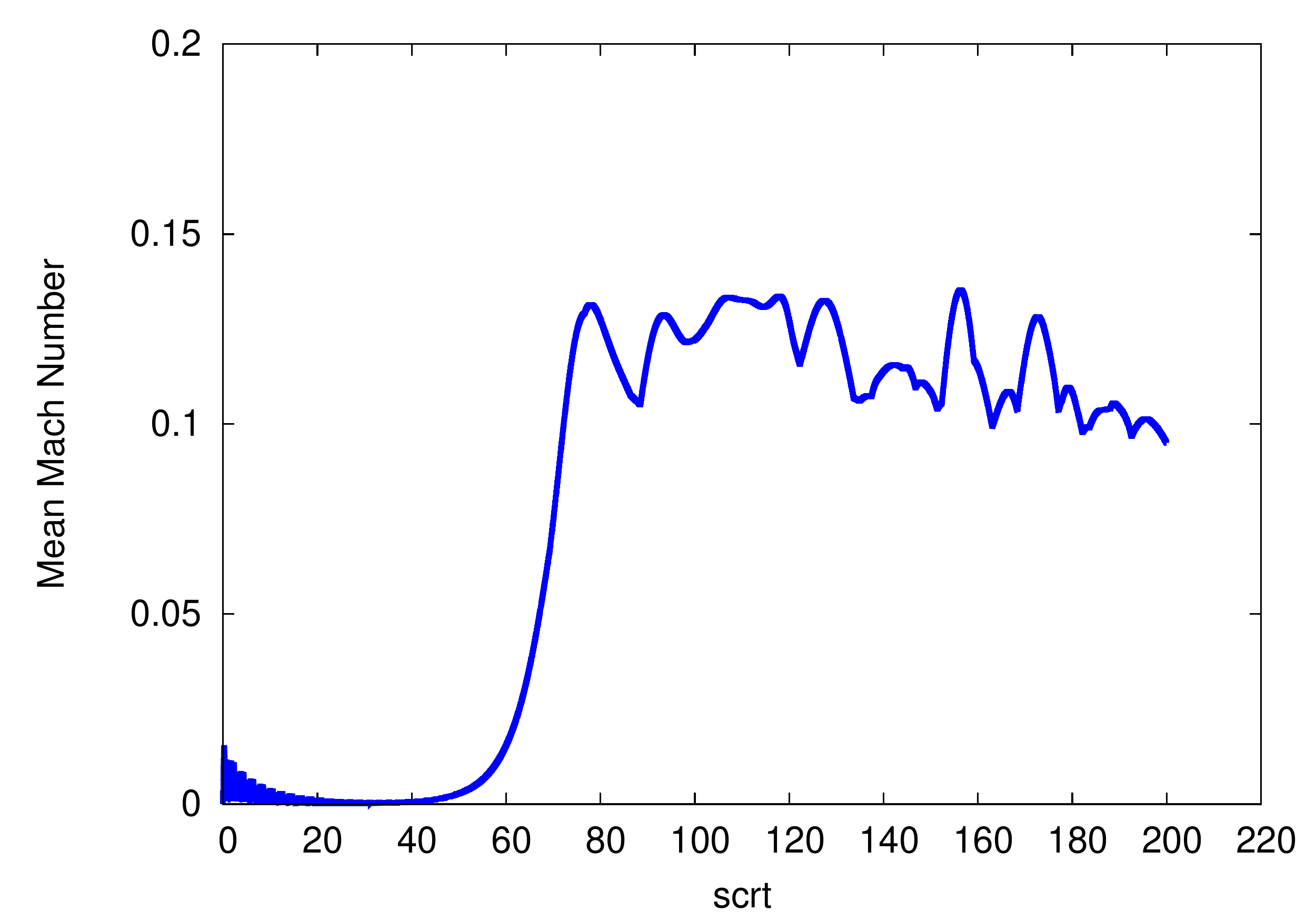}
\caption{Simulation of double-diffusive convection with parameters $Pr=1.0$, $Ra = 160 000$, $Le = 0.01$ and $R_{\rho} = 1.3$. The spatial resolution is 600 $\times$ 400 grid points. Note that in this setting diffusion poses an even greater restriction on the timestep than sound waves.} 
\label{fig:DS5}
\end{figure}

However, in some settings within our parameter range, the timestep restriction due to diffusive processes is also severe and in some cases it is even more restrictive
than the acoustic timestep restriction (see Figure \ref{fig:DS5}). Note that the resolution is somewhat lower in this case (only 3 grid points resolve the diffusive boundary layer), but for the present purpose this is sufficient, especially since at high resolution the diffusive timescales are even more restrictive.   

Following the idea of integrating terms describing processes acting on short timescales implicitly, we have combined total-variation diminishing implicit-explicit (TVD IMEX) Runge-Kutta methods with our solver. This combination treats the diffusive terms implicitly and therefore alleviates the timestep restrictions due to diffusive processes. In \cite{imex2012} TVD IMEX Runge-Kutta methods are analysed in detail and the fruitful junction of our solver and those methods is demonstrated.

%The combination of the numerical method presented in this work with total-variation-diminishing implicit-explicit (TVD IMEX) Runge-Kutta methods provides
%a solution to this difficulty by integration the diffusive terms of the Navier -Stokes equations implicitly and thereby alleviating the corresponding
%timestep restrictions. The merits of TVD IMEX Runge-Kutta methods are exposed in \cite{imex}.

\section*{Acknowledgements}
We acknowledge financial support from the Austrian Science fund (FWF), projects P20973 and P21742. 
We thank the referees for their helpful comments. 

\bibliographystyle{elsart-num-sort}
\bibliography{NH-Bibliography}

\end{document}